\documentclass[12pt]{article}
\usepackage{bezier}
\usepackage{amssymb}
\usepackage{epsfig}

\newtheorem{lemma}{Lemma}[section]

\newtheorem{definition}{Definition}[section]

\newcommand{\nc}{\newcommand}
\nc{\be}{\begin{equation}}
\nc{\ee}{\end{equation}}
\nc{\bea}{\begin{eqnarray}}
\nc{\eea}{\end{eqnarray}}

\nc{\eqn}[1]{{(\ref{#1})}}
\nc{\cA}{{\cal A}}
\nc{\cB}{{\cal B}}
\nc{\cC}{{\cal C}}
\nc{\cD}{{\cal D}}
\nc{\cE}{{\cal E}}
\nc{\cF}{{\cal F}}
\nc{\cG}{{\cal G}}
\nc{\cH}{{\cal H}}
\nc{\cI}{{\cal I}}
\nc{\cJ}{{\cal J}}
\nc{\cK}{{\cal K}}
\nc{\cL}{{\cal L}}
\nc{\cM}{{\cal M}}
\nc{\cN}{{\cal N}}
\nc{\cO}{{\cal O}}
\nc{\cP}{{\cal P}}
\nc{\cQ}{{\cal Q}}
\nc{\cR}{{\cal R}}
\nc{\cS}{{\cal S}}
\nc{\cT}{{\cal T}}
\nc{\cU}{{\cal U}}
\nc{\cV}{{\cal V}}
\nc{\cW}{{\cal W}}
\nc{\cX}{{\cal X}}
\nc{\cY}{{\cal Y}}
\nc{\cZ}{{\cal Z}}
\nc{\simo}[1]{{\stackrel{#1}{\simeq}}}
\nc{\geqo}[1]{{\stackrel{#1}{\geq}}}
\nc{\geo}[1]{{\stackrel{#1}{>}}}
\nc{\guo}[1]{{\stackrel{#1}{\succ}}}

\nc{\rbo}{\raisebox}
\nc{\RR} {\rangle \! \rangle}
\nc{\LL} {\langle \! \langle}
\nc{\rmi}[1]{{\mbox{\small #1}}}
\nc{\eq}{eq.~}
\nc{\nr}[1]{(\ref{#1})}
\nc{\ul}{\underline}
\nc{\mc}{\multicolumn}
\nc{\todo}[1]{\par\noindent{\bf $\rightarrow$ #1}}

\nc{\cu}{{\cal u}}

\title{
  \begin{flushright} {\small HD-THEP-96-07} \end{flushright}
\vskip 2cm
Critical Phenomena with \\ Linked Cluster Expansions \\
in a Finite Volume}

\author{Hildegard~Meyer-Ortmanns\thanks{Email address
h.meyer-ortmanns@thphys.uni-heidelberg.de}\\
  and \\
 Thomas~Reisz\thanks{Heisenberg fellow, Email address
reisz@thphys.uni-heidelberg.de}
         \\ \\Institut
        f\"ur Theoretische Physik,\\
        Universit\"at Heidelberg, \\
        Philosophenweg 16, \\
        D-69120 Heidelberg, Germany}

\begin{document}

\maketitle

\begin{abstract}
Linked cluster expansions are generalized from an infinite to a finite
volume. They are performed to 20th order in the expansion parameter to
approach the critical region from the symmetric phase.
A new criterion is proposed to distinguish 1st from 2nd order
transitions within a finite size scaling analysis. The criterion
applies also to other methods for investigating the phase structure
such as Monte Carlo simulations. Our computational tools are illustrated
at the example of scalar O(N) models with four and six-point couplings
for $N=1$ and $N=4$ in three dimensions.
It is shown
how to localize the tricritical line in these models. We indicate some
further applications of our methods to the electroweak transition as well as
to models for superconductivity.
\end{abstract}

%
%

\newpage

\section{Introduction}

The phase structure of models for strong and electroweak interactions has been
a topic of intensive research in the past. In spite of numerous investigations
some central questions are still open. To these belong the nature
of the chiral/
deconfinement transition in QCD for physical values of the current quark
masses and the strength of the electroweak transition for the physical (so far
unknown) Higgs mass. In both realms one has to account for nonperturbative
coupling regions. Thus it is natural to choose the lattice regularized
version of these theories to study their phase structure. Most applications
are performed with Monte Carlo simulations, which are an appropriate tool
to study the critical region.
Monte Carlo simulations are restricted to a finite volume. Thorough
extrapolations to the infinite volume limit from a finite size scaling analysis
is in general expensive and sometimes impracticable for lattice sizes which
are realistic for QCD or for the standard model \cite{hilde}.

Convergent expansions such as
Linked Cluster or Hopping Parameter Expansions (HPEs)
provide an analytic alternative to
Monte Carlo simulations. They may also serve as a convenient
supplement to numerical calculations.
Originally
they have been developed in the {\it infinite} volume.
In contrast
to generic perturbation theory about  noninteracting
fields, HPEs are convergent power series expansions
about completely disordered lattice systems.
Under certain conditions their convergence radius can be
directly related to the location of the physical singularity.
Hence,
similarly to Monte Carlo
simulations, HPEs can be applied to the phase transition (critical) region, if
the order in the hopping parameter $\kappa$
is just high enough. Thus the transition
region is accessible from the high temperature (symmetric) phase.

Hopping parameter expansions have a long tradition in statistical physics
(\cite{Wortis,ID,guttmann1} and references therein).
Their generalization and application to particle
physics have been pioneered by L\"uscher and Weisz \cite{LW1}.
L\"uscher and Weisz studied a
lattice $\Phi^{4}$-theory close to its continuum limit
in four dimensions \cite{LW2,LW3}.
Recently, the HPE has been
generalized to
field theories at finite temperature \cite{thomas1}.
The generalization is twofold. First of all one has to implement a
toroidal symmetry in one direction of finite extension, say $L_0$.
In this context the temperature $T$ is given by
$T=L_0^{-1}$ in lattice units.
Second, the highest computed order
in the expansion parameter has to be increased, because the toroidal
(temperature) effect on the critical coupling is rather small.
Typically, the critical hopping parameter
$\kappa_c$ changes only by a few percent even on a
$4\times\infty^3$ lattice compared to the
$\infty^4$ lattice.
The graphs of the expansion can only
"feel the temperature", if they are able to wind around the torus
in the temperature direction. In general, the largest possible winding
number should be larger than one to induce a measurable effect.
In \cite{thomas2} the 18th order has been used to determine
the critical behaviour of the finite temperature $\Phi^4$ models
with O(N) symmetry.
Meanwhile, the 20th order of this expansion is available
for 2-point susceptibilitites, see below.

In this paper we extend the HPE to a finite volume, i.e. to a lattice
with toroidal symmetry in all directions.
We propose criteria to
distinguish first from second order transitions (and crossover phenomena) both
in an infinite and in a finite volume.
It is the fate of  power series expansions that one cannot work {\it at}
the singularity $\kappa_c$, one can only come close to it, the
closer,
the higher the order in the expansion.
Thus we need a criterion that works slightly below $\kappa_c$.
As such a criterion we propose a so called monotony criterion which is
based on the specific volume dependence of truncated correlation
functions close to but not at the transition point.
The criterion includes both order parameter susceptibilities
and other singular response functions such as the specific heat.
Decrease or increase with the volume identifies
first or second order transitions, respectively.
Although the monotony criterion has been developed in the framework of
HPEs, it is not restricted to this case.
It can be used in other methods as well,
in particular in  Monte Carlo simulations.

As a second application of the HPE in a finite volume we calculate an
effective potential up to 16th order in the hopping
parameter.
The shape of the effective potential further characterizes the type of
transition. The coexistence of distinct minima at the critical
point
provides another possibility to calculate $\kappa_c$ in a finite volume.

The criteria will be applied to scalar O(N)-models
with $\Phi^{4}$ and $\Phi^{6}$ self-\-inter\-actions in three
dimensions. These models allow for various
first and second order transition regions in the bare
coupling constant space.
For fixed couplings the phase transitions will be considered as a function
of $\kappa$. The parameter $\kappa$ may be identified with an inverse
temperature $1/T$ of a classical system with the same action in three
dimensions. Thus we will sometimes replace $\kappa_c$ by $T_c^{-1}$. In
connection with a field theory in four spacetime dimensions the three
dimensional model may be considered as an effective description of the
four dimensional model at finite temperature, arising in a process of
dimensional reduction. In a four dimensional theory at finite temperature
one should distinguish between $\kappa$ and $T^{-1}$.

The scalar O(N)-models contain a number of interesting special cases.
If the four-and
six-point couplings are sent to infinity
in an approriate way,
we obtain "diluted" $O(N)$ models, i.e. Heisenberg models with
additional
occupation number variables, for $N=1$ we have a diluted Ising model.
The case of $N=4$
and pure quartic selfinteraction
is assumed to share the universality class with QCD in the limit of
two massless flavours. It also corresponds to the scalar sector of
the electroweak standard model.
A $\Phi^4 +
\Phi^{6}$-theory exhibits a tricritical point (line) for a fixed (varying)
six-point coupling. Such a tricritical point is observed in a liquid mixture
of $He^3$/$He^4$.
Recently it has been also proposed as candidate for representing the
universality class of {\it tricritical QCD} \cite{wilczek,rajagopal}.
(Tricritical QCD means
QCD with vanishing up and down quark masses and a strange quark mass
which takes a
critical value, at which the chiral transition changes its order.) We
indicate how to localize the tricritical line in a $\Phi^4 +
\Phi^6$-theory with our methods.

The outline of
the paper is as follows. In section 2.1 we summarize the
main results
for HPEs from \cite{thomas1}. It basically serves to fix the notation.
We then extend
the HPE in an infinite volume to a graphical expansion in a {\it finite}
volume (section 2.2).
In section 3 we give two criteria to distinguish 1st from 2nd order
transitions: a precise formulation of the {\it monotony criterion}
(section 3.1), and an effective potential
evaluated in the HPE in a
finite volume (section 3.2). In section 4 we apply these criteria to
three-dimensional scalar O(N) models
with renormalizable interactions. To get a first estimate
on the phase structure
in bare coupling parameter space, we study the large
coupling limit
by a saddle point integration. Another
estimate for the location at finite couplings is
obtained from a hopping-mean-field analysis. This approximation amounts to a
tree level evaluation of the HPE
(Section 4.1).
After this preliminary study of the phase structure  we
present a more detailed investigation
by means of the HPE.
In the infinite volume limit, plateaus
of critical exponents as obtained from the linked cluster series
are proposed as criteria
to identify the various universality classes of the critical region
of the theory (Section 4.2). In Section 4.3 we discuss the
finite volume behaviour of various quantities.
The shift in volume of the critical coupling $\kappa_c$,
defined here as the radius of convergence, is compared to
the scaling behaviour which is expected for the shift of the maximum of
the order parameter susceptibility.
The monotony criterion and the effective potential are evaluated for
points both in the first and second order coupling
region. Finally we show how to locate the tricritical line.
In section 5 we summarize our
results and give an outlook to further physical applications.

%
%


\section{\label{lce.0}Hopping parameter expansions for the critical
region}

\subsection{General framework}

Linked cluster expansions
provide a convenient tool for both numerical and analytic
studies of lattice field theories. The typical expansion
parameters are the coupling strengths between
fields at different lattice sites
\cite{Wortis,ID}.
In contrast to saddle point expansions which are at most asymptotically
convergent, series resulting from HPE are absolutely convergent
for sufficiently small couplings \cite{andreas1}. In this sense
they can be viewed as generalized high temperature expansions.
If in addition the sign
of susceptibility series is uniform,
the radius of convergence identifies the phase transition, i.e. the
critical temperature.

In order to extract quantitative information on the critical behaviour
one has to get sufficiently close to the critical point.
The price to be paid is a computation to high orders. The realization
of such expansions by convenient algorithms
with the aid of computers has been pioneered by
L\"uscher and Weisz \cite{LW1}.
Recently, progress in various ways has been made to extend
the length of strong coupling series \cite{thomas1,butera,campost}.
Normally, these expansions are set up in an infinite
volume.
In \cite{thomas1}, the techniques have been further developed
in such a way that the expansions can be reliably applied to
lattices of non-trivial topology.
In particular, it turned out that the highest order
in the expansion had to be further
increased
for measuring
effects from topology.
The improved techniques have been applied to scalar O(N) models with
quartic interaction on
4-dimensional finite temperature lattices \cite{thomas2}.
The critical exponents could be shown to agree with the critical
indices of
the corresponding (dimensionally reduced) 3-dimensional models.
\noindent

In the following we summarize the main formulas
from~\cite{LW1,thomas1} to fix the notation and to set up the
expansion scheme that later will be generalized to a finite volume.
We consider a $D$-dimensional hypercubic lattice $\Lambda=
\times_{i=0}^{D-1}{\mathbb{Z}}/L_i$,
with $L_i\in{\mathbb{N}}$ an even number or with $L_i=\infty$. Periodic
boundary conditions are imposed for each finite $L_i$.
The restriction to even $L_i$ leads to a
considerable reduction of the number of contributing
graphs because it implies that each loop must have an
even number of lines.
The class of models we discuss are described by the partition function
\bea \label{lce.1}
   Z(J,v) & = & \int \prod_{x\in\Lambda} d^N\Phi(x) \;
   \exp{(  \frac{1}{2}
    \sum_{x\not=y\in\Lambda}\sum_{a,b=1}^N\Phi_a(x)v_{ab}(x,y)
      \Phi_b(y) )} \nonumber \\
   &&\cdot \exp{(- \sum_x \stackrel{\circ}{S}(\Phi(x))
      + \sum_{x\in\Lambda} \sum_{a=1}^N J_a(x) \Phi_a(x) )},
\eea
where $\Phi$ denotes a real, $N$-component scalar field, $J$ are
external sources, $v_{ab}(x,y)$ denote the hopping couplings.
The ultralocal part of the action $\stackrel{\circ}{S}$,
which depends only on one lattice site, is chosen to be
$O(N)$ invariant. It should guarantee the stability of the partition function
\eqn{lce.1} for sufficiently small $v(x,y)$.
Throughout this paper we consider as an example
the action of a $\Phi^4 + \Phi^6$-theory
\be \label{lce.3}
   \stackrel{\circ}{S}(\Phi) \; = \; \Phi^2 + \lambda (\Phi^2-1)^2
   + \sigma (\Phi^2-1)^3,
\ee
which
exhibits a phase structure with both first and second order
transitions. We emphasize, however, that the general techniques are not
restricted to this case.

Fields at different lattice sites interact with the
hopping coupling
$v_{ab}(x,y)$. For the case of nearest neighbour interactions,
it reduces to
\be \label{lce.4}
   v_{ab}(x,y) \; = \; \left\{
   \begin{array}{r@{\qquad ,\quad} l }
    2\kappa\;\delta_{a,b}\; & {\rm x,y \; nearest\; neighbour} \\
    0 & {\rm otherwise},
   \end{array} \right.
\ee
where $\kappa$ is the so called hopping parameter.
The nearest neighbour property should be understood modulo the torus lengths.
Henceforth we consider only  nearest neighbour interactions.

The generating functional of connected correlation functions
is given by
\bea
  W(J,v) & = & \ln{ Z(J,v) }, \nonumber \\
 \label{lce.5}
  W_{a_1\ldots a_{2n}}^{(2n)} (x_1,\ldots, x_{2n}) & = &
   <\Phi_{a_1}(x_1) \cdots \Phi_{a_{2n}}(x_{2n}) >^c \\
   & = & \left.
   \frac{\partial^{2n}}{\partial J_{a_1}(x_1) \cdots
   \partial J_{a_{2n}}(x_{2n})}
    W(J,v) \right\vert_{J=0}. \nonumber
\eea
In the following a major role is played by the connected 2-point
function and the corresponding susceptibility $\chi_2$ and moments $\mu_2$,
defined according to
\bea
  \delta_{a,b}\; \chi_2 & = &
   \sum_x < \Phi_a(x)\Phi_b(0) >^c , \nonumber \\
 \label{lce.10}
  \delta_{a,b}\; \mu_2 & = &
   \sum_x \left( \sum_{i=0}^{D-1} x_i^2 \right)
    < \Phi_a(x)\Phi_b(0) >^c .
\eea
In field theory, it is convenient to define the renormalized coupling
constants via
the vertex functional
\bea \label{lce.gamma}
  \Gamma (M) & = & W(J) - \sum_{x\in\Lambda} J(x)\cdot M(x) \nonumber \\
   & = & \sum_{n\geq 0} \frac{1}{2n!} \sum_{a_1,\ldots, a_{2n}}
  \Gamma_{a_1\ldots a_{2n}}^{(2n)}(x_1,\ldots, x_{2n}) \;
   M_{a_1}(x_1) \cdots M_{a_{2n}}(x_{2n}), \\
  M_a(x) & = & \frac{\partial W}{\partial J_a(x)}\; ,\quad
   a=1,\ldots ,n .\nonumber
\eea
The standard definitions of the renormalized mass $m_R$
(as inverse correlation length) and the
wave function renormalization constant $Z_R$
are
\be \label{lce.gamma2}
  \widetilde\Gamma_{ab}^{(2)}(p,-p) \; = \; - \frac{1}{Z_R} \,
   ( m_R^2 + p^2 + O(p^4) ) \; \delta_{a,b} \quad
   {\rm as} \;p \to 0 ,
\ee
where $\;\widetilde{ }\;$ denotes the Fourier
transform. Eq.~\eqn{lce.gamma2}
implies that
\be \label{lce.renorm}
  m_R^2 = 2 D \frac{\chi_2}{\mu_2} \; , \;
  Z_R = 2 D \frac{\chi_2^2}{\mu_2} .
\ee
The critical exponents $\gamma,\nu,\eta$ are defined by the
leading singular behaviour at the critical point $\kappa_c$,
\bea
  \ln{\chi_2} & \simeq & -\gamma \ln{(\kappa_c-\kappa)}, \nonumber \\
  \label{lce.expo}
  \ln{m_R^2} & \simeq & 2\nu \ln{(\kappa_c-\kappa)}\; , \quad
 \mbox{as}\; \kappa\nearrow\kappa_c , \\
  \ln{Z_R} & \simeq & \nu\eta \ln{(\kappa_c-\kappa)}, \nonumber
\eea
such that
$\nu\eta=2\nu-\gamma$.

If the interaction part \eqn{lce.4}
of the action is switched off, i.e.~$v=0$,
$S(\Phi,v=0)=\sum_x\stackrel{\circ}{S}(\Phi(x))$, the partition
function factorizes, and in turn
$W(J,v=0)=\sum_x\stackrel{\circ}{W}(J(x))$. In particular,
\be \label{lce.13}
  W_{a_1\ldots a_{2n}}^{(2n)} (x_1,\ldots, x_{2n}) \; = \; \left\{
   \begin{array}{r@{\qquad ,\quad} l }
    \frac{\stackrel{\circ}{v}_{2n}^c}{(2n-1)!!} C_{2n}(a_1,\ldots , a_{2n})
       & {\rm for \; x_1=x_2=\cdots=x_{2n}} \\
    0 & {\rm otherwise}
   \end{array} \right.
\ee
with
\be \label{lce.14}
  \stackrel{\circ}{v}_{2n}^c = \left.
   \frac{\partial^{2n}}{\partial J_1^{2n}}\; \stackrel{\circ}{W}(J)
   \right\vert_{J=0},
\ee
and $C_{2n}$ totally symmetric coefficients in $a_i, i=1...2n$.

In practice, the vertex couplings
$\stackrel{\circ}{v}_{2n}^c$ are obtained from the relation
\bea
    \stackrel{\circ}{W}(J) & = & \sum_{n\geq 1} \frac{1}{(2n)!}
    \stackrel{\circ}{v}_{2n}^c (J^2)^n \nonumber \\
  \label{lce.vcc}
  & = & \ln{(1+\sum_{n\geq 1}  \frac{1}{(2n)!}
    \stackrel{\circ}{v}_{2n} (J^2)^n )},
\eea
with
\be
     \stackrel{\circ}{v}_{2n} = \frac{ \int d^N\Phi
       \Phi_1^{2n} \exp{(-\stackrel{\circ}{S}(\Phi))} }
       { \int d^N\Phi
        \exp{(-\stackrel{\circ}{S}(\Phi))} },
\ee
or, alternatively, recursively from Dyson-Schwinger equations.

The linked cluster expansion for $W$ is the
Taylor expansion with respect to
$v(x,y)$ about this decoupled case,
\be \label{lce.15}
  W(J,v) = \left. \left( \exp{\sum_{x,y} \sum_{a,b} v_{ab}(x,y)
   \frac{\partial}{\partial \widehat v_{ab}(x,y)}} \right) W(J,\widehat v)
   \right\vert_{\widehat v =0}.
\ee
The corresponding expansions of correlation functions are
obtained from \eqn{lce.15} by \eqn{lce.5}.
Susceptibilities become power series in $\kappa$
with a nonvanishing radius of convergence.

The management of such an expansion is conveniently done by means of
a graph theoretical device. Correlation functions are represented
as a sum over equivalence classes of graphs,
each class being endowed with an appropriate weight.
In order to make high orders in the expansion feasible it is necessary
to introduce more restricted
subclasses of graphs such as 1-particle irreducible (1PI) graphs,
1-vertex irreducible graphs, and renormalized moments. The correlations
are then represented in terms of the latter two. For further details
we refer to \cite{LW1,thomas1}.

The weight of each graph decomposes into a product of its
(inverse) symmetry number, the O(N)-group factor, and the
lattice embedding number.
It is only the latter one
that depends on the topology of the particular lattice which is involved.
In the next section we outline the modifications of the
embedding numbers due to a finite
volume.

\subsection{Extension to the torus}

Let us consider a correlation function, such as in \eqn{lce.10},
on a $D$-dimensional lattice of size $L_0 \times L_1
\times ... \times L_{D-1}$ with periodic boundary conditions.
Except for a trivial volume factor,
the embedding number
$I_\Gamma (L_0,\ldots,L_{D-1})$ of a connected graph $\Gamma$
counts the number of  possible
ways $\Gamma$ can be embedded on the lattice.
Embedding means a mapping of every vertex $v$ of $\Gamma$ onto a lattice
site $x(v)=(x_0,\ldots, x_{D-1})(v)$
consistent with the topology of $\Gamma$.
Every two vertices have to be mapped to nearest neighbour lattice sites
if they are neighboured vertices of $\Gamma$, i.e. if they are connected
by at least one line. Selflines do not exist. Otherwise the linked
cluster expansion does not impose any exclusion constraints. In particular,
an arbitrary number of vertices can occupy the same lattice
site.

It is most convenient to rearrange the computation of embedding
numbers in terms of random walks \cite{LW1}.
Towards this end, the set of vertices of $\Gamma$ is devided in
the disjoint sets of internal 2-vertices and their complement.
A vertex $v$ is called internal 2-vertex, if it has no external line
attached, and there are precisely two neighboured vertices of
$v$ in $\Gamma$.
All internal 2-vertices can be reorganized into so-called
2-chains between the remaining vertices in an obvious way.
Every 2-chain $c$ has an initial vertex $i_c$ and a final vertex
$f_c$, possibly identical, and it has a length $l_c\geq 1$,
where $l_c-1$ denotes the number of internal 2-vertices of $c$.
Here, for convenience, we include $l_c=1$, in which case $c$ just
implies the nearest neighbour constraint on $i_c$ and $f_c$.
On the lattice infinite in all directions
the embedding number
is then written as
\be \label{lcetorus.1}
   I_\Gamma(\infty^D) = \sum_{ ( x(v) ) }^{\qquad\prime} \; \prod_c
   \cN_{ x(i_c) \to x(f_c) }^{l_c,D} ( \infty^D ).
\ee
The sum runs over all placements of vertices $v$ that are
not internal 2-vertices, with $x(v_0)$ kept fixed for some arbitrary
vertex $v_0$ to account for the trivial entropy factor.
The product runs over all 2-chains of $\Gamma$.
$\cN_{x\to y}^{l,D}(\infty^D)$ denotes the
number of free random walks of length $l$ from lattice site
$x$ to $y$.
Closed analytic expressions can be
given for $\cN^{l,D}_{x\to y}$.
We notice that
\be \label{lcetorus.2}
   \cN_{x\to y}^{l,D}(\infty^{D}) \not= 0 \quad\mbox{only if}\quad
   l - \sum_{i=0}^{D-1} | x_i - y_i | \; \geq \; 0
   \quad \mbox{even}.
\ee

In the finite volume with periodic boundary conditions
the topology  modifies \eqn{lcetorus.1} at two places.
First, the sites $x(v)$ are now restricted to a cube of size
$L_0 \times L_1\times ... \times L_{D-1}$, and the nearest
neighbour constraint, implicit in every 2-chain $c$ of length
$l_c=1$ holds modulo the torus lengths.
Second, the number of random walks
$\cN_{x\to y}^{l_c,D}(\infty^D)$ is replaced by
\be \label{lcetorus.3}
  \cN_{x\to y}^{l_c,D}(L_0,\ldots, L_{D-1}) \; = \;
   \sum_{\mu_0,\ldots,\mu_{D-1}\in\mathbb{Z}}
  \cN_{x\to y+\mu\cdot L}^{l_c,D}(\infty^D),
\ee
where
\[
  \mu \cdot L \; = \; \sum_{i=0}^{D-1} \mu_i L_i .
\]
The sum in \eqn{lcetorus.3} accounts for additional random walks which
arise from the possible winding around the torus.
Due to \eqn{lcetorus.2}, the sum in
\eqn{lcetorus.3} is finite.

%
%


\section{\label{mon.title}Finite size scaling analysis with HPE}

In this section we discuss two criteria in the finite volume
to determine the order of a phase transition. Although the criteria
are developed for linked cluster expansions, their application is not
restricted to series representations.

The question arises, why one is interested in linked cluster expansion
on a torus, since
the expansion are more easily obtained in
the infinite volume.
Data of the critical region such as
the critical temperature are successfully
extracted from the asymptotic high order
behaviour of the coefficients of susceptibility series.
The typical precision here
is within 4-5 digits or even better.

In general the symmetry of the model alone does not determine
 the properties of a transition.
For instance, there may be more than one universality class,
corresponding to different ranges in the space of bare
actions.
As was pointed out in \cite{thomas2}, one should look for plateaus
of critical exponents to distinguish between them.
Problems arise
close to the boundary of two such domains. "Smearing effects"
occur due to the truncation of the series.
Both universality domains will influence the
coefficients, the more the lower the order.
This does not pose a  problem, as long as the domains are sufficiently large
and the boundary of the domains extends over a negligible coupling range.

Whereas the location of the phase transition can be determined
very precisely from the infinite volume series,
its order is usually a more intricate question, in particular,
if the transition is weakly first order.
Criteria to distinguish 1st and 2nd order transitions
can be conveniently worked out in the finite volume.
Thus, in our case, the finite size effects will be utilized rather
than suppressed as artifacts of the finite volume.
\noindent

A finite size scaling analysis for second order transitions can be
based on a renormalization group approach, see e.g. \cite{barbour}. The inverse
linear size $L^{-1}$ of the system is put on an equal footing with
other scaling fields like the temperature or an external field. The
analysis results in a prediction of the leading scaling behaviour of a
susceptibility $\chi$
about the critical temperature $T_c$ according to
\be \label{monintro.1}
\chi (t,L) \simeq {|t|}^{-\gamma} \; P(L/\xi_\infty(t))
\ee
for sufficiently small $t$ and large $L$,
where $t$ is the reduced temperature $(T-T_c)/T_c$,
and $\gamma$ is the critical exponent
characterizing the divergence at $t=0$.
The amplitude $P$
depends only on $L$ measured in units of
the infinite volume correlation length $\xi_\infty$.
Further properties of $P$ ensure that the height
of the peak of the susceptibility at $T_c(L)$
scales according to $\chi(T_c(L),L) \simeq L^{(\gamma/\nu)}$,
the width $\sigma(L)$ of
the critical region shrinks with $L^{-1/\nu}$, and
$T_c(L)$ is shifted compared to $T_c$ according to $T_c(L)-T_c$ $\simeq
L^{-1/\nu}$. Here $\nu$ denotes the critical exponent
of the correlation length.

For a generic first order transition an analogous derivation
of the scaling behaviour from first principles
is missing in general. The ratio $L/\xi_\infty (t)$ is no
longer a distinguished scaling variable. In the thermodynamic limit,
as $T$ approaches $T_c$, the correlation stays finite and model
dependent. The rounding and shifting of thermodynamic singularities
are normally described in a phenomenological approach \cite{binder}
which is based on Monte Carlo results. The height of the peak
of the susceptibility at $T_c(L)$
is expected to scale with $L^D$, both
the width $\sigma(L)$ and the shift
in $T_c(L)-T_c$ are expected to scale with $L^{-D}$ as $L\to\infty$,
where $D$ denotes the
space(time) dimension.
A more fundamental finite size scaling theory exists for a class
of spin models that allow for a particular polymer expansion
of the partition function \cite{borgs}.
Whereas the predictions of the finite size scaling of the height
of the peak and the width of
the scaling region
reproduce the above mentioned behaviour,
the shift of the location of the peak is derived  to be
$T_c(L) - T_c$  $\simeq L^{-2D}$.

It turns out that our series representations in the hopping parameter
$\kappa$ for the susceptibilities
$\chi$ cannot be uniquely extrapolated to the critical $\kappa_c$
(corresponding to $T_c^{-1}$ as explained in the introduction)
to the end that the peak
and width of $\chi$ confirm the expected scaling behaviour. However,
the specific behaviour of $\chi$ close to $\kappa_c$ is conclusive
enough for distinguishing regions of 1st and 2nd order transitions, as
we will show below,
without any need for an extrapolation in $\kappa$ to $\kappa_c$.
In addition, the scaling of
$\kappa_c(L)$, defined as the radius of convergence
in the finite volume, follows the form  expected (by analogy)
for the shift of the location of the peak from \cite{borgs}.
The scaling behaviour holds for the
$\mathbb{Z}_2$ model, but also for the $\Phi^4+\Phi^6$-models with
four components, which are not covered by the analysis of
\cite{borgs}, cf. section IV.3.

\subsection{The monotony criterion}

For a certain interval of the scaling region response
functions  with
a nonanalytic behaviour in the infinite volume limit show different
monotony behaviour for 1st and 2nd order transitions.
Examples for such functions are the specific heat and order parameter
susceptibilities. They are
increasing in volume in a certain neighbourhood of $T_c$ for 2nd order
transitions, and decreasing for 1st order transitions for some
range in the scaling region, which will be specified below. Since we are
not aware of any discussion in the literature, although the underlying
idea is rather simple, we will describe the behaviour in some detail.
For definiteness we fix the notation in terms of order parameter
susceptibilities.

{}From the standard finite size scaling analysis one knows that
\be \label{monintro.2}
  \chi (t,L=\infty ) \; < \; \infty \quad \mbox{as} \; t\to 0
\ee
for a 1st order transition with a possible discontinuity, whereas
\be \label{monintro.3}
  \chi (t,L=\infty) \simeq \cA |t|^{-\gamma}
\ee
for a 2nd order transition with critical exponent $\gamma>0$.
By definition, regular contributions to $\chi$ may be neglected in the
scaling region.
On the other hand, $\chi(0,L)$ diverges in both cases as
$L$ approaches infinity. More precisely,
at $T_c$ , $\chi$ has a "$\delta$-function" or power law type of
singularity for a 1st or 2nd order transition in the thermodynamic limit,
respectively.
It is this difference
 that is responsible for
the different monotony properties for $t\not=0$ in the finite volume.
If $t\not=0$ is small, to a given lattice size $L_s < \infty$,
not too small, one can always
find a second size $L_l$ with
$L_l>L_s$ such that
\bea \label{monintro.4}
  \chi(t, L\geq L_l ) > \chi(t, L_s)
   \quad & & \mbox{for 2nd order} , \\  \label{monintro.5}
  \chi(t, L\geq L_l ) < \chi(t, L_s)
   \quad & & \mbox{for 1st order} ,
\eea
that is, $\chi$ is increasing or decreasing in volume.
In the series representation of $\chi$ we can set $L_l=\infty$,
so that \eqn{monintro.4} and \eqn{monintro.5}
give strong criteria in the whole scaling region where we
can use the series.
If the volume cannot be made infinite, as in Monte Carlo
simulations, one is confined to $L\leq L_0$ for some $L_0$.
Eq. \eqn{monintro.5} then still holds except for
a small neighbourhood of $T_c$ , where the susceptibility is increasing in
volume as in the 2nd order case. The width of this neighbourhood is
rapidly decreasing with increasing $L_0$.

In the following we make these statements more precise.
Let
$t$ denote the scaling field, i.e. $t=(T-T_c)/T_c$, $v$  the
inverse volume or an
appropriate power of it. The infinite volume limit is obtained
as $v\to 0$ from above. The transition range is then given by
small $t$ and $v$.
Let
\be \label{mon.h2}
   H^2 := \{ (t,v)\in{\bf R}^2 \; \vert
     \; v\geq 0 \}
\ee
denote a half plane, $\cU\subseteq H^2$ an open neighbourhood
of $0\in H^2$ and $\cU^* = \cU\setminus\{0\}$.

We discuss the case of a first order transition first.
The typical behaviour of a susceptibility close to the
transition is described by the following

\begin{definition}\label{mondef1o}
For any $\omega >0$ we define $\Psi_1^\omega(\cU)$ as the set of
real valued continuous functions
$\chi: \cU^* \to {\bf R} $
with the following properties.
\begin{description}
\item{1.} $\chi\in C^1(\cU^*\setminus\{({\bf R},0)\})$,
that is, $\chi$ is once continuously
differentiable for $v\not=0$.
\item{2.} For all nonzero $t$ there is $\nu_t>0$ such that for all
$v < |t|/\nu_t$
\[
    | \; \chi(t,v) \; | \leq \omega .
\]
\item{3.} With appropriate positive constants $c,K_1,K_2$ and $\epsilon$
we have in $\cU^*$ for $v\not= 0$
\bea
  | \; \chi (0,v) - \frac{c}{v} \; | & < & \frac{K_1}{v^{1-\epsilon}} ,
  \nonumber \\
  | \; \frac{\partial}{\partial t} \chi(t,v) \; |
  & < & \frac{K_2}{v^2} . \nonumber
\eea
\end{description}
\end{definition}

As an example, consider the following representation of the magnetic
susceptibility in the volume $L^D$
\be \label{monchi21o}
   \chi_2(T ,L) = c L^D \; \exp{(-f L^{2D}
    (T - T_c + d L^{-2D})^2 )}
    + \eta(T,L),
\ee
with $c,d,e$ constants and $\eta(\cdot,L)$ analytic for $L<\infty$,
locally uniformly convergent as $L\nearrow\infty$
(so that $\eta(\cdot,\infty)$ is analytic).
With $v=L^{-D}$, $t=(T-T_c)/T_c$, it is straight forward to show that
\[
    \chi(t,v) := \chi_2(T,L)
\]
belongs to $\Psi_1^\omega(\cU)$ for some $\omega$.

More generally, every function $\chi:\cU^*\to{\bf R}$ of the form
\be \label{monf1o}
   \chi(t,v) \; = \; \frac{1}{v} \; f(\frac{t}{v})
   + \widetilde\chi(t,v)
\ee
belongs to $\Psi_1^\omega(\cU)$ with appropriate $\omega>0$,
if the following conditions are satisfied.
\begin{description}
\item{1a.} $\widetilde\chi(t,v)\in C^1(\cU^*)$,
\item{1b.} $\widetilde\chi(t,v)$ together with its (first) partial
derivatives are uniformly bounded in $\cU^*$.
\item{2a.} $f\in C^1({\bf R})$ is a nonnegative function with $f(0)>0$,
\item{2b.} $\lim_{x\to\pm\infty} |x|^{1+\epsilon} f(x) = 0$
for some $\epsilon>0$,
\item{2c.} $(d/dx)f(x)$ is uniformly bounded on ${\bf R}$.
\end{description}

Any such function has the property that it "approaches $\delta$"
locally, i.e.
\[
    \lim_{\epsilon\to 0+} \lim_{v\to 0+}
    \int_{-\epsilon}^{\epsilon} dt \; \chi(t,v) \; > \; 0
\]
and is finite. In this case the limits do not commute.
An explicit example for such a  function is
\[
     f(x) \; = \; (\frac{c}{\pi})^{1/2} \exp{(-cx^2)} \; , \; c>0
\]

with normalization
\[
    \int_{-\infty}^{\infty} dx \; f(x) = 1.
\]

After these preliminaries
we now state the following volume behaviour.

\begin{lemma}\label{monlem1o}
Let $\omega>0$, $\chi \in \Psi_1^\omega(\cU)$.
There are $\delta$, $\epsilon > 0$, and for every $t\not=0$ there is
$\nu_t>0$ such that in $\cU^*$ for all $w,v,t$ with
$v < \delta$ and $\nu_t w < |t| < \epsilon  v$,
\[
   \chi(t,v) \; > \; | \chi(t,w) | .
\]
\end{lemma}

In particular the lemma holds for $w=0$, i.e.
\[
   \chi(t,v) \; > \; |\chi(t,0)|.
\]
This means that the susceptibility in that part of the transition region
where $|t|/v < \epsilon$
is larger than in the infinite volume limit where $w=0$.
\noindent

{\sl Proof:}
Let $\chi\in\Psi_1^\omega(\cU)$, $\omega>0$.
Differentiability implies that
\[
   \chi(t,v) = \chi(0,v) + t \int_0^1 ds
   \left. \frac{\partial}{\partial\eta} \chi(\eta,v)
   \right|_{\eta = st}.
\]
With appropriate $c_0,K>0$ we have
\[
   \chi(0,v) \geq \frac{c_0}{v}
\]
and
\[
   \left| \frac{\partial}{\partial\eta} \chi(\eta,v)
   \right| < \frac{K}{v^2}
\]
in $\cU^*$. Hence
\[
   \chi(t,v) \geq \frac{c_0}{v} - \frac{|t| K}{v^2}.
\]
Furthermore, for every $t\not=0$ there is $\nu_t>0$ such that
\[
   | \chi(t,w) | \leq \omega
\]
for all $w < |t|/\nu_t$.
Finally, we choose $\epsilon=c_0/(2K)$ and
$\delta=c_0/(4\omega)$ and get for
$v<\delta$ and $\nu_t\; w < |t| < \epsilon\; v$
\[
   \chi(t,v) \geq \frac{c_0}{v} - \frac{\epsilon K}{v}
   = \frac{c_0}{2v} > |\chi(t,w)|,
\]
thus it follows the lemma.
$\qquad\square$

Now we come to the second order transition.
In contrast to the first order case, at a second order transition
the order parameter susceptibility can be divergent
in the infinite volume limit
as the critical temperature is approached. This is described by

\begin{definition}\label{mondef2o}
For any $\gamma >0$ we denote by $\Psi_2^\gamma(\cU)$ the set of functions
$\chi: \cU^* \to {\bf R}$ that are continuous and satisfy
the following conditions.
\begin{description}
\item{1.} There are constants $\cA$, $K$, $\epsilon>0$ such that in $\cU^*$
\[
    | \; \chi(t,0) \; - \; \cA\; |t|^{-\gamma} \; |
    \leq \; K \; |t|^{-\gamma+\epsilon} .
\]
Furthermore, with appropriate $\nu$, $\cC>0$, we have whenever
$|t| > \nu\; v$,
\[
   \chi(t,v) \; \geq \; \cC \; \chi(t,0).
\]
\item{2.} There are constants $\eta$, $\cB>0$ such that for $|t|<\eta\; v$
\[
   | \; \chi(t,v) \; | \; < \; \cB \; v^{-\gamma}.
\]
\end{description}
\end{definition}

The specific  property for a 2nd order transition
that the singular part of the
free energy density behaves as a generalized homogeneous
function implies for the susceptibility in a volume $L^D$
a typical form like
\be
   \chi_2(T ,L) = |T-T_c|^{-\gamma} \; Q( (T-T_c) L^{1/\nu} )
    + \eta(T,L),
\ee
with some $\gamma >0$.
Here $\eta(\cdot,L)$ has similar analyticity properties as in \eqn{monchi21o}
above, $\nu>0$ is the critical exponent of the correlation length
\be
   \xi \sim | T-T_c |^{-\nu},
\ee
$Q$ is continuous and behaves as
\bea
   \lim_{x\to 0} |x|^{-\gamma} Q(x) & = & K > 0, \nonumber \\
    \lim_{x\to\pm\infty} Q(x) & = & C > 0 .
\eea
The first equation expresses
the absence of a nonanalyticity of $\chi_2$ for finite $L$, the
second one its presence in the infinite volume case.
With $t=(T-T_c)/T_c$, $v=L^{-1/\nu}$ and
\[
   \chi(t,v) \; := \; \chi_2(T,L)
\]
we see that $\chi$ belongs to $\Psi_2^\gamma(\cU)$ for some $\gamma$.

More generally, every function $\chi:\cU^*\to{\bf R}$ of the form
\be
   \chi(t,v) \; = \; \frac{1}{v^\gamma} \; f(\frac{t}{v})
   + \widetilde\chi(t,v)
\ee
with $\gamma >0$
belongs to some $\Psi_2^\gamma(\cU)$, if the following conditions
are satisfied.
\begin{description}
\item{1.} $\widetilde\chi(t,v)\in C^1(\cU^*)$.
\item{2a.} $f\in C^1({\bf R})$, and $f(0)>0$,
\item{2b.} $\lim_{x\to\pm\infty} |x|^{\gamma} f(x) = C$
for some finite $C>0$.
\end{description}

Compared to the first order case \eqn{monf1o} the
essential difference comes from property (2b).

As an example,
\[
    \chi(t,v) \; = \; (t^2+v^2)^{-(m/2)} \;, \; m>0,
\]
belongs to the class
$\Psi_2^{m}$.

For these functions, we have in contrast to Lemma \eqn{monlem1o}

\begin{lemma}\label{monlem2o}
Let $\gamma>0$ and $\chi\in\Psi_2^\gamma(\cU)$.
There are constants $\nu$, $\epsilon>0$, $\nu<\epsilon$,
such that for all $t,v,w$ with $\nu w < |t| < \epsilon v$
\[
   | \; \chi(t,v) \; | \; \leq \; \chi(t,w) .
\]
\end{lemma}

The inequality is always true if $w=0$, i.e.
the susceptibility is always smaller than in the infinite
volume limit as long as we are in the critical
region $|t|/v < \epsilon$.

{\sl Proof:}
Let $\chi\in\Psi_2^\gamma(\cU)$, $\gamma>0$.
There are numbers $\cC,\cD,\nu>0$ such that in $\cU^*$ for
$|t| > \nu\; w$
\[
   \chi(t,w) \; > \; \cC \; \chi(t,0) \; > \; \cD \; |t|^{-\gamma}.
\]
Furthermore, there are $\eta,\cB>0$ such that for
$|t| < \eta\; v$
\[
   | \chi(t,v) | \; < \; \cB \; v^{-\gamma}.
\]
We choose $\epsilon=\min{(\eta, (\cD/\cB)^{1/\gamma})}$
and get for
$\nu\; w < |t| < \epsilon\; v$
\[
   | \chi(t,v) | \; < \;  \cB \; v^{-\gamma}
   \; < \; \cB \left(\frac{\epsilon}{|t|}\right)^\gamma
   \; < \; \frac{\cB\epsilon^\gamma}{\cD} \; \chi(t,w)
   \; < \; \chi(t,w).
\]
This proves the lemma.
$\qquad\square$

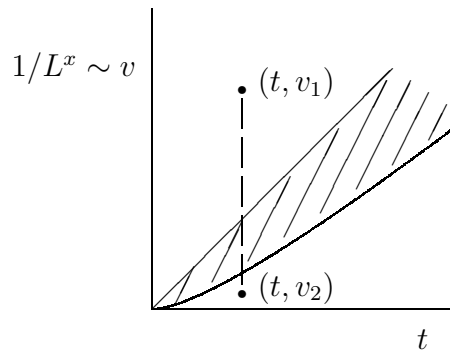
\begin{figure}[htb]
\caption{\label{monfig1} $(t,v)$-plane for
 susceptibilities $\chi(t,v)>0$ in the vicinity of a phase transition at
$(t=0,v=0)$, $t$ denotes the scaling field
$t=(T-T_c)/T_c$, $v$ is inverse to some power of the volume $L^x$
with some $x>0$.
For a 1st~order transition, $\chi(t,v_1)>\chi(t,v_2)$,
whereas for a 2nd~order transition
$\chi(t,v_1)<\chi(t,v_2)$. For the shaded part no prediction
is made.}

\begin{center}
\setlength{\unitlength}{0.8cm}
\begin{picture}(10.0,5.0)


\put(2.0,0.0){
\setlength{\unitlength}{0.8cm}
\begin{picture}(10.0,5.0)

\put(0.0,0.0){\line(1,0){5.0}}
\put(-1.8,4.0){\makebox(1.0,0){$1/L^x\sim v$}}
\put(0.0,0.0){\line(0,1){5.0}}
\put(4.0,-0.5){\makebox(1.0,0){$t$}}

\qbezier(0.0,0.0)(1.0,0.0)(5.0,3.0)
\put(0.0,0.0){\line(1,1){4.0}}

\put(1.5,3.65){\circle*{0.12}}
\put(1.9,3.75){\makebox(1.0,0){$(t,v_1)$}}
\put(1.5,0.25){\circle*{0.12}}
\put(1.9,0.35){\makebox(1.0,0){$(t,v_2)$}}
\put(1.5,0.4){\line(0,1){0.5}}
\put(1.5,1.05){\line(0,1){0.5}}
\put(1.5,1.7){\line(0,1){0.5}}
\put(1.5,2.35){\line(0,1){0.5}}
\put(1.5,3.0){\line(0,1){0.5}}
\put(0.4,0.1){\line(1,2){0.3}}
\put(1.0,0.45){\line(1,2){0.5}}
\put(1.6,0.8){\line(1,2){0.7}}
\put(2.2,1.2){\line(1,2){0.9}}
\put(2.8,1.6){\line(1,2){1.1}}
\put(3.4,2.0){\line(1,2){0.9}}
\put(4.0,2.4){\line(1,2){0.6}}
\put(4.6,2.8){\line(1,2){0.3}}

\end{picture}
}

\end{picture}
\end{center}

\end{figure}


\subsection{The effective potential}

Let us briefly discuss  another method to determine the nature
of a transition that will be of use later on. For definiteness
we come back to the $N$-component scalar model as described
in Section~\ref{lce.0}.
A possible way to define an effective potential is by
\be \label{monconn.2}
    V \cdot V_{\rm eff} (M) \; = \;
    - \left. \Gamma(M) \right|_{M=\rm const},
\ee
where $V$ denotes the volume, and $\Gamma(M)$ is defined by
\eqn{lce.gamma}.
In the symmetric phase and in the infinite volume limit, $V_{\rm eff}$
has to be convex.
In practice, the right hand side of \eqn{monconn.2}
is obtained as an expansion about $M=0$.
In the linked cluster expansion the coefficients can be
expressed in terms of 1PI susceptibilities
$\chi_n^{\rm 1PI}$.
They are obtained as series representation in $\kappa$ of the
truncated susceptibilities
by keeping only those graphs that are 1PI \cite{thomas2}.
Up to a constant we obtain
\bea \label{mon.effpot}
  V_{\rm eff}(M) & = &
  \frac{1}{2}\; \frac{1-4D\kappa \chi_2^{\rm 1PI}}{\chi_2^{\rm 1PI}}
   M^2
   - \frac{1}{4!} \; \frac{\chi_4^{\rm 1PI}}{(\chi_2^{\rm 1PI})^4}
   (M^2)^2 \\
  &  & -
  \frac{1}{6!} \; \frac{1}{(\chi_2^{\rm 1PI})^6}
  \; \left( \chi_6^{\rm 1PI} - \frac{10 (\chi_4^{\rm 1PI})^2}{\chi_2^{\rm 1PI}}
  \right) (M^2)^3
  + O(M^8). \nonumber
\eea
Any nonconvex shape of $V_{\rm eff}$ in the infinite volume limit
must be an artifact of the approximation scheme. In a finite volume,
however, a nonconvex shape in the symmetric phase
signals a first order transition, whereas a convex shape is compatible
with a second order transition. An estimate of the critical coupling is then
 obtained by a root of the coefficient of
$M^{2}$ in the 2nd order case, and by degenerate values of $V_{\rm eff}$
at the trivial and nontrivial minima in the 1st order case. In our
applications (see Section 4) we have calculated the
1PI-susceptibilities $\chi^{\rm 1PI}_{2n}$ up to 16th order in $\kappa$ in
a finite volume for $n=1,2,3$.

%
%
\section{Applications to scalar O(N) models with $\Phi^{4}$ and
$\Phi^{6}$-point couplings}

In this section we apply the methods discussed in the
previous sections to the three-dimensional O(N) symmetric
scalar model with $N=1$ and $N=4$.
The model is described on a lattice
$\Lambda$ by the partition function
\be \label{exa.part}
   Z(\kappa,\lambda,\sigma) = \int \prod_{x\in\Lambda} d^N\Phi(x) \;
   \exp{( 2\kappa
    \sum_{x,y\; {\rm NN}} \Phi(x) \cdot
      \Phi(y)
    - \sum_x\stackrel{\circ}{S}(\Phi(x),\lambda,\sigma)
          )},
\ee
where the first sum of the exponential runs over unordered
pairs of nearest neighbour lattice sites, and the ultralocal part
$\stackrel{\circ}{S}$ is given by
\be \label{exa.action}
   \stackrel{\circ}{S}(\Phi,\lambda,\sigma)
    \; = \; \Phi^2 + \lambda (\Phi^2-1)^2
   + \sigma (\Phi^2-1)^3
\ee
with $\sigma >0$ or $\sigma=0$ and $\lambda\geq 0$.
In contrast to the pure quartic interaction which only
admits 2nd order transitions, the action \eqn{exa.action}
allows a richer
phase structure with regions of 1st and 2nd order transitions due to
the additional $\Phi^{6}$ interaction. The case of $\lambda=3\sigma$
corresponds to a pure $\Phi^{6}$-theory, whereas $\lambda<3\sigma$
implies a negative quartic coupling.

\subsection{\label{exa.sub1}Preliminaries}

To get a first estimate of the phase structure we consider the case
of large couplings $\lambda$ and $\sigma$. For finite coupling
constants we invoke a hopping mean field analysis.

{\it The large coupling limit.}
To study the limit of large couplings, we set
$\lambda = \alpha \sigma$ and send $\sigma$ to infinity
with  $\alpha$ and $\kappa$ kept finite and fixed.
This limit is discussed for more general contact terms
in Appendix \ref{app1.0}.
The discussion is based on a saddle point integration.
As a result we obtain the following behaviour in dependence
on $\alpha$.
For $\alpha>1$ and $\alpha<-3$
we obtain O(N) Heisenberg models (Ising model for $N=1$).
The range of $-3<\alpha<1$ leads to complete disordering with no
phase transition at any finite $\kappa$.
The cases of $\alpha=-1$ and $\alpha=3$ are peculiar. The
resulting actions describe "diluted" O(N) models, with particular
values of the couplings. If the large coupling limit is performed term
by term in
 the HPE series it can be shown that at least for $N\geq 2$
the resulting actions again belong to the unversality class of
O(N) Heisenberg models.

{\it A hopping-mean-field analysis.}
To get a first estimate of the phase structure at {\it finite
couplings}, it
is instructive to start with a mean field analysis.
Together with the convexity of the exponential and the positivity
of the measure this ansatz leads to a complete factorization of the
partition function.
The hopping mean field estimate for the free energy is then derived
as follows.
Let us define
$\stackrel{\circ}{x}$ by
\[
   \exp{\stackrel{\circ}{x}\!(H)} \; = \;
   \int d^N\Phi \;
   \exp{( - \sum_x\stackrel{\circ}{S}(\Phi,\lambda,\sigma)
          + H \cdot \Phi
          )},
\]
and the expectation value $< F >_H$ of an observable $F$ according to
\bea
   < F(\Phi) >_H \; & = & \;
    \exp{(-|\Lambda| \stackrel{\circ}{x}\!(H) )} \nonumber \\
    && \cdot \;
    \int \prod_{x\in\Lambda} \left(
   d^N\Phi(x) \; \exp{( - \sum_x\stackrel{\circ}{S}(\Phi)
          + H \cdot \Phi )} \right)
    F(\Phi) . \nonumber
\eea
Here $|\Lambda|$ denotes the lattice volume, and $H\in\mathbb{R}^N$
is an auxiliary field.
Note that $\stackrel{\circ}{x}$ defined in this way agrees with
$\stackrel{\circ}{W}$ as introduced in Sect.~\ref{lce.0}.
In particular, for every integer $n$,
\[
  \stackrel{\circ}{v}_{2n}^c(\lambda,\sigma) = \left.
   \frac{\partial^{2n}}{\partial H_1^{2n}}\; \stackrel{\circ}{x}\!(H)
   \right\vert_{H=0}
\]
(cp.~to \eqn{lce.14}).
For simplicity, where no confusion
can arise,
we only indicate the dependence on $H$.
We get
\bea
   Z(\kappa,\lambda,\sigma) & = &
   <  \exp{( 2\kappa
    \sum_{x,y\; {\rm NN}} \Phi(x) \cdot
      \Phi(y)
          )} >_H  \nonumber \\
   & \geq &   \exp{ < 2\kappa
    \sum_{x,y\; {\rm NN}} \Phi(x) \cdot
      \Phi(y) >_H  }  \nonumber \\
   & = & \exp{(-|\Lambda| \; \overline{f}(H) )}\nonumber
\eea
with
\[
  \overline{f}(H) \; = \;
  - \left( \stackrel{\circ}{x}\!(H) + 6\kappa
    (\nabla_H \!\!\stackrel{\circ}{x}\!(H))^2
    - H \cdot  \nabla_H \stackrel{\circ}{x}\!(H) \right).
\]
An upper bound on the true free
energy density $f$ is thus given by
\be
  f \; \leq \; \inf_H \overline{f}(H).
\ee
A vanishing
$H_0=0$ is always a solution of the corresponding mean field equation
\[
   \partial_H\overline{f} (H_0) \; = \;
   - \partial_H^2\!\!\stackrel{\circ}{x}\!(H_0) \left(
   12\kappa \partial_H\!\!\stackrel{\circ}{x}\!(H_0) - H_0 \right) = 0,
\]
where the derivative is in the direction of $H$,
and $H_0=0$ is a local minimum of $\overline{f}$, if
\be \label{exa.stable}
   \partial_H^2\overline{f} \; = \;
   \partial_H^2\!\!\stackrel{\circ}{x}\!(H_0) \left(
   1 - 12\kappa \;\partial_H^2\!\!\stackrel{\circ}{x}\!(0) \right)  > 0.
\ee
In particular, for small hopping parameter $\kappa$ the model
is always in the symmetric phase.

This form of mean field analysis is identical to the tree level of
the hopping parameter expansion.
It can be easily shown that the Lebowitz inequality
\[
   \partial_H^3\!\!\stackrel{\circ}{x}\!(H) < 0
   \;\;\;\mbox{for all}\; H>0
\]
together with $\partial_H^2\!\!\stackrel{\circ}{x}\!(H)>0$
for all $H$
ensures that $H_0=0$ is the absolute minimum of
$\overline{f}$. The Lebowitz inequality holds in any case for
$\lambda\geq 3\sigma$.
Equality in \eqn{exa.stable} along with
\be \label{exa.quart}
   \partial_H^4\overline{f}(0) > 0
\ee
locates a 2nd order phase transition to the spontaneoulsy
broken phase at
\[
   \kappa_c(\lambda,\sigma) \; = \; \frac{1}
    {12\stackrel{\circ}{v}_{2}^c(\lambda,\sigma)}.
\]
{\it Tricritical points} are identified by an
equality in \eqn{exa.quart} and
\be \label{exa.six}
   \partial_H^6\overline{f}(0) > 0,
\ee
i.e.
\[
   \stackrel{\circ}{v}_{4}^c(\lambda,\sigma) = 0 \;\mbox{and}\;
   \stackrel{\circ}{v}_{6}^c(\lambda,\sigma) < 0.
\]

We notice that these conditions imply a vanishing two and
four-point coupling in the effective potential $V_{eff}$, Eq.\eqn{mon.effpot},
evaluated to tree level in the HPE, i.e. in the
hopping mean field approximation. Thus the criterion for
tricriticality
reduces to the familiar one. If we had used the {\it classical} potential
\eqn{exa.action} instead,
this would lead to a location of the tricritical line in the bare
coupling constant space at $\lambda=3\sigma$.


\begin{table}[htb]
\caption{\label{exa.tri1} Mean field tricritical line for
the O(4) model
on the three-dimensional hypercubic lattice.
The line is defined by $\stackrel{\circ}{v}_{4}^c = 0$ and
$\stackrel{\circ}{v}_{6}^c<0$,
$\overline\sigma$ is defined by
$\overline\sigma = \stackrel{\circ}{v}_{6}^c
/(5(\stackrel{\circ}{v}_{2}^c)^3).$
}
\vspace{0.5cm}

\begin{center}
\begin{tabular}{|r|r|r|r|}
\hline \hline
$\alpha\quad$ & $\lambda=\alpha\sigma$ &
$\sigma\quad$ & $\overline\sigma\qquad$   \\
[0.5ex] \hline
 1/1.05  & 128.199   & 134.609    & -0.7245305       \\
 1/1.1   & 53.040   & 58.344    & -0.6875585      \\
 1/1.2   & 21.822     & 26.286    & -0.5920738      \\
 1/1.3   & 13.252   & 17.228     & -0.5193738      \\
 1/1.5   & 7.315      & 10.973    & -0.4451270     \\
 1/2.0   & 3.401    & 6.802    & -0.3839411     \\
 1/5.0   & 0.788    & 3.941    & -0.3407590     \\
 1/10.0  & 0.344    & 3.494   & -0.3345469     \\
 1/100.0 & 0.031   & 3.081    & -0.3306763     \\ [0.5ex] \hline
 -1/10.0 & -0.273   & 2.731    & -0.3284753     \\
 -1/1.1  & -1.391   & 1.530    & -0.3519023     \\
 -1.0    & -1.472   & 1.472   & -0.3581768    \\
 -2.0    & -2.948    & 1.474     & -0.5106579     \\
 -3/1.2  & -7.561   & 3.024  & -0.6827047    \\
 -3/1.1  & -17.678 & 6.482  & -0.7259408     \\\hline
\end{tabular}

\end{center}

\end{table}

In Table 1 we have listed some results
for the O(4)-model on the location of the tricritical
line in the $(\lambda,\sigma)$-space for several values of $\alpha$.

The {\it tricritical exponents} in hopping-mean-field establish the results
of the most  naive mean field analysis with the effective potential
replaced by the classical potential. They are
$\alpha=1/2, \beta=1/4, \gamma=1, \delta=5, \nu=1/2$ .
{}From the Ginzburg criterion one may expect that the only chance where
a mean field type of analysis may lead to reliable
predictions of the singular behaviour in {\it three} dimensions is
{\it at} tricriticality.
In fact, the susceptibility comes out as volume
independent along the tricritical line (cf. Section 4.3 below)
when it is determined by the
HPE analysis. A mean field analysis is
volume independent by construction.

\subsection{Infinite volume analysis at finite couplings}

So far we have studied the phase structure in the large coupling limit
and in a mean field analysis for finite couplings. Next we utilize
the linked cluster expansions for a more thorough study.
Susceptibilities are represented as convergent power series in the
hopping parameter, such as the $2n$-point functions
\be \label{exa.chi}
   \chi_{2n}(\kappa,\lambda,\sigma) =
   \sum_{\mu\geq 0} a_\mu^{(2n)}(\lambda,\sigma) \kappa^\mu ,
\ee
and similarly for weighted correlations.
These series have been computed to 20th order in $\kappa$ for $n=1$,
to 18th order for $n=2$, and to 16th order for $n=3$, both in a finite
and infinite volume. In the infinite volume, the coefficients of the
series we have explicitly calculated are of equal sign for each
series. Under the assumption that this behaviour continues to all orders
in $\kappa$,
the radius of convergence
$\kappa_c(\lambda,\sigma)$ is identified with the singularity
closest to the origin
on the positive real axis, hence with
the physical singularity at the phase
transition, independently of the
order of the transition.

Well developed methods are known to obtain critical data from the
high order coefficients of  high temperature series
\cite{guttmann1,gaunt}.
The critical point $\kappa_c(\lambda,\sigma)$
is identified by the ratio criterion, applied to the
coefficients $a_\mu^{(2n)}(\lambda,\sigma)$.
The best choice is the 2-point susceptibility, because its series
is available to the highest order.
The obliged regression towards large $\mu$ is done according to
\be \label{exa.rmu}
  r_\mu := \left| \frac{a_\mu^{(2)}}{a_{\mu-1}^{(2)}} \right| =
  \frac{1}{\kappa_c} \left( 1 + \frac{c_1}{\mu^{\omega_1}}
  + O(\mu^{-\omega_2}) \right)
\ee
with $\omega_2>\omega_1>0$, $c_1$ as fit parameter
chosen according to the best $\chi^2/df$ fit.
This procedure is eventually  supplemented by a shift of the
weak antiferromagnetic
singularity at $-\kappa_c$ to $-\infty$, an improved estimator fit
and other known techniques like Pade` methods.
We know that $\omega_1=1$ for a leading pole or branch point singularity
on the real axis, i.e. for
\[
   \chi_2 \simeq \cA \left( \kappa_c - \kappa \right)^{-\gamma} \;
   \; \mbox{as} \; \kappa \nearrow \kappa_c
\]
with $\gamma>0$ and $\gamma\not= 1$,
and $\omega_1>1$ for $\gamma=1$.
For a 2nd order transition, an alternative way to determine
the critical point is given by the smallest real solution of
\[
   12 \kappa_c \chi_2^{\rm 1PI}(\kappa_c) \; = \; 1 ,
\]
as proposed in \cite{thomas2}. This condition is equivalent to the
identification
of the phase transition as a zero of the quadratic coefficient of
the effective potential, cf. \eqn{mon.effpot}.
It turns out to be the most convenient way to determine
the radius of convergence leading to the
highest precision in $\kappa_c$. The lowest precision
obtained in this way lies within 4-5 digits.
Once $\kappa_c$ is determined we obtain the critical exponent $\gamma$
from
\be \label{exa.gamma}
   1 + \mu (\kappa_c r_\mu - 1 ) \; = \;
   \gamma + \frac{c_2}{\mu^{\omega_3}} + o(\mu^{-\omega_3}),
\ee
with $\omega_3$ and $c_2$ as fit parameters.
In a similar way, the critical exponent $\nu$ is obtained by replacing
the series of $\chi_2$ by that of $m_R^{-2}$, cf. \eqn{lce.renorm}.

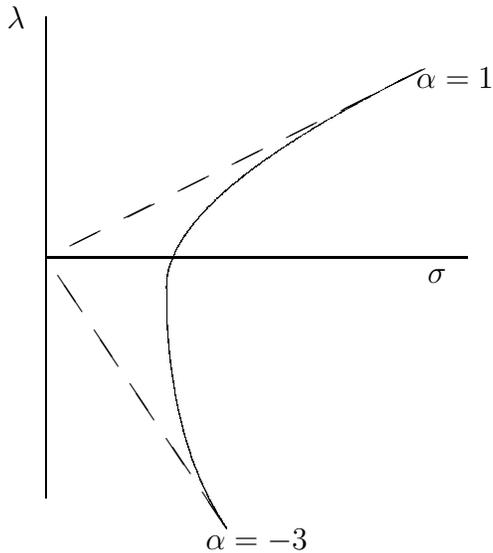
\begin{figure}[htb]
\caption{\label{exaphase} Qualitative plot of the phase
structure of O(N) lattice models in 3 dimensions.
The dashed curves give the lines $\lambda=\alpha\sigma$
with $\alpha =1$ and $\alpha=-3$.
The solid curve represents the tricritical line
$\lambda_t(\sigma)$.
To the left of it the phase transition is of 2nd order,
to the right of it 1st order.
}

\begin{center}
\setlength{\unitlength}{0.8cm}
\begin{picture}(10.0,10.0)


\put(2.0,0.0){
\setlength{\unitlength}{0.8cm}
\begin{picture}(10.0,10.0)

{\thinlines
\put(0.0,5.0){\line(1,0){7.0}}
\put(6.0,4.7){\makebox(1.0,0){$\sigma$}}
}
{\thinlines
\put(0.0,5.0){\line(0,1){4.0}}
\put(0.0,5.0){\line(0,-1){4.0}}
\put(-1.0,9.0){\makebox(1.0,0){$\lambda$}}
}


{\thinlines
\put(0.4,5.2){\line(2,1){0.5}}
\put(1.3,5.65){\line(2,1){0.5}}
\put(2.2,6.1){\line(2,1){0.5}}
\put(3.1,6.55){\line(2,1){0.5}}
\put(4.0,7.0){\line(2,1){0.5}}
\put(4.9,7.45){\line(2,1){0.5}}
\put(5.8,7.9){\line(2,1){0.5}}
}

\put(6.3,8.0){\makebox(1.0,0){$\alpha =1$}}


{\thinlines
\put(0.2,4.7){\line(2,-3){0.4}}
\put(0.8,3.8){\line(2,-3){0.4}}
\put(1.4,2.9){\line(2,-3){0.4}}
\put(2.0,2.0){\line(2,-3){0.4}}
\put(2.6,1.1){\line(2,-3){0.4}}
}

\put(3.0,0.3){\makebox(1.0,0){$\alpha =-3$}}


{\linethickness{0.1pt}
\qbezier(6.0,8.0)(2.0,6.0)(2.0,4.5)
\qbezier(3.0,0.5)(2.0,2.0)(2.0,4.5)
}

\end{picture}
}

\end{picture}
\end{center}

\end{figure}


A measurement of critical exponents like $\gamma,\nu,\eta$ leads to
a qualitative plot of the phase structure in
the$(\lambda,\sigma)$-halfplane
as shown in
Fig.~\ref{exaphase}.
The solid line represents the boundary $\lambda=\lambda_t(\sigma)$
between the 2nd and 1st order region.
To the left of this tricritical line the phase transition is of
2nd order. Here,
except for the origin $\lambda=\sigma=0$, we obtain one universality
class for every $N$ with plateaus of critical exponents,
with values of the $N$-component Heisenberg model
(cf.~e.g.~\cite{thomas2} for a recent list of those exponents).
It is remarkable that this range considerably extends the "Lebowitz domain"
$\lambda\geq 3\sigma$, where the action is convex and therefore
$\gamma$ is not less than 1 \cite{froehlich}.
In particular, it includes the full range of $\lambda<-3\sigma$.
In passing we mention that for $\lambda\geq 0$ the
presence of a small nonvanishing $\Phi^6$ interaction,
that is $\sigma>0$,
considerably accelerates the convergence of the high
temperature series compared to the case $\sigma=0$, i.e.
the non-universal remainder of \eqn{exa.rmu}
becomes smaller.

Except for the values of the critical exponents, the phase structure
qualitatvely confirms the mean field analysis
of the last subsection.
As the tricritical line
$\lambda=\lambda_t(\sigma)$ is approached,
the exponents $\gamma$ and $\nu$ drop continuously from their
 values in the Heisenberg models to the Gaussian values, where the
tricritical line is crossed, and
further to zero.
The smooth interpolation between the different exponents is an
artifact of
the truncation of the
power series expansions at high, but finite order in $\kappa$.
It was already pointed out in \cite{thomas2} that various
universality classes lead to smearing effects
at finite order due to an "interference" of various universality
domains. The most pronounced plateau structure is obtained
for $\nu\eta$, the critical exponent
of the wave function renormalization constant $Z_R$ \eqn{lce.expo}.

\begin{figure}[htb]
\caption{\label{nueta} Critical exponent
$\nu\eta$ of the wave function renormalization constant $Z_R$
for the 3d O(4) model, obtained from the 20th order
susceptibility series for various $\sigma$
along the ray $\lambda = (1/2) \sigma$.
The tricritical point is at $\sigma_t\simeq 9.0$.
The mean field estimate gives $\sigma_t=6.8$.
}

\setlength{\unitlength}{0.8cm}
\begin{picture}(12.0,12.0)

\epsfig{file=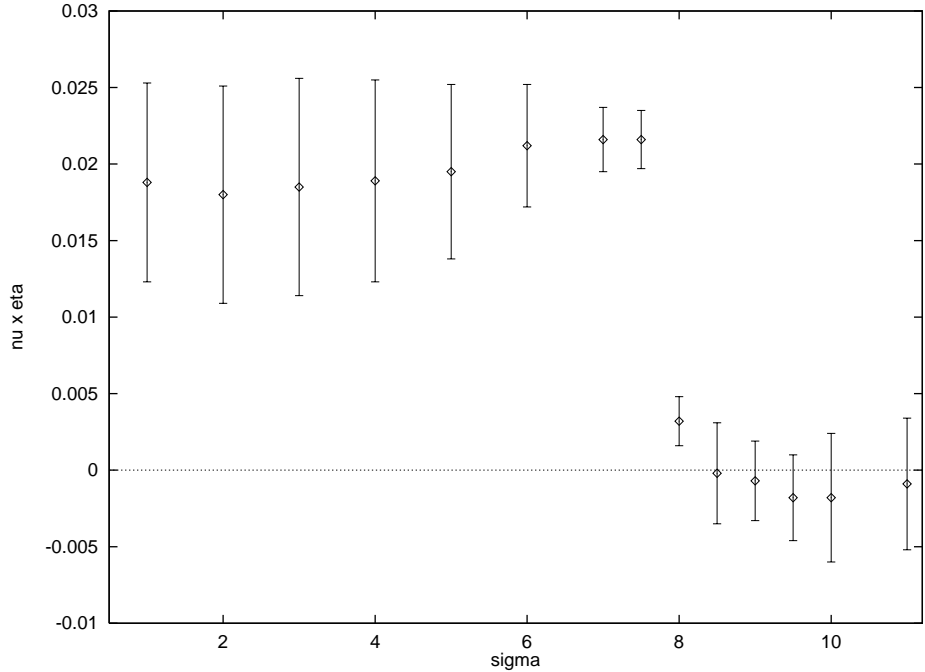}

\end{picture}

\end{figure}


In Fig.~\ref{nueta} we show the results on the exponent $\nu\eta$ for the
$O(4)$ model, obtained from the 20th order susceptibility series of
$\chi_2$ and $\mu_2$, \eqn{lce.10},
for various $\sigma$ along the ray $\lambda=(1/2)\sigma$.
There is a well established plateau at the left part of the plot
corresponding to the universality class of the Heisenberg model.
The plateau at the right part
is compatible with a range of 1st order transitions, all exponents
$\gamma, \nu, \eta$ vanish within the error bars. Hence they are
compatible with a finite correlation length at $\kappa_c$. The
stability of the extrapolated convergence radius $\kappa_c$ under a
variation of the truncation of the series suggests that there is
really a 1st order
transition rather than a mere crossover phenomenon.
We would like to identify the left boundary of this
plateau as the tricritical point. This gives us an estimate
of about $\sigma_t\simeq 9.0$. The indicated errors in Fig.~\ref{nueta}
are obtained
as discussed in connection with \eqn{exa.gamma}.
The smearing effect does not allow for a more precise location of the
tricritical point.
To get a clearer identification of
the 1st order transition region and a better localization of the
tricritical point it is natural to perform
a finite size scaling
analysis, which is the topic of the next section.

\subsection{Results of the finite size scaling analysis}

In Sect.~\ref{mon.title} we have formulated monotony criteria
for response functions $\chi$ in the scaling region.
In absolute value, increase in volume implies a 2nd order transition,
decrease in volume a 1st order one.
The response functions
have to be calculated at some $\widetilde\kappa$ close to but not at
the critical point $\kappa_c$ for two volumes. The value of
$\widetilde\kappa$ should be chosen sufficiently close to the
transition point to satisfy the conditions of the monotony criterion,
and sufficiently apart from $\kappa_c$ to allow a use of the
truncated series
representation. A choice of $\widetilde\kappa=0.98\kappa_c(\lambda,\sigma)$
fulfills both restrictions.

The volumes should be sufficiently large in lattice units to guarantee
the applicability of the finite size scaling ansatz for $\chi$. Beyond
this generic condition the following restrictions arise from the
monotony criterion. The smaller one of the two volumes should satisfy
$L^x |\widetilde\kappa - \kappa_c| \lesssim 1$
with $x=1/\nu$ or $x=3$, which implies an upper bound on $L$ for given
$\widetilde\kappa$. In practice we have chosen this $L$ between $4$ and
$12$. In the context of the HPE the larger one of the two volumes
may be set to infinity. The advantage of this choice is that
in the 1st order case,
decrease in volume holds all over the
scaling region. In Monte Carlo simulations the
second volume is necessarily finite. In this case $L$ should be large
enough so that $\widetilde\kappa$ lies outside the small neighbourhood
of $\kappa_c$ where $\chi$ is increasing in the 1st order case as well.
The critical point $\kappa_c(\lambda,\sigma)$, which enters the
inequalities on $L$ and $\widetilde\kappa$, is determined
as the radius of convergence in the infinite volume
as described above.

We use the 2-point
susceptibility , because its series is available to the highest order
in $\kappa$. For given couplings $\lambda$ and $\sigma$
\be \label{fss.chi2}
   \chi_2^{(M)}(\widetilde\kappa,\lambda,\sigma;L) =
   \sum_{\mu=0}^M a_\mu^{(2)}(\lambda,\sigma;L) \; \widetilde\kappa^\mu
\ee
denotes the 2-point susceptibility truncated at order $M$. With
\be \label{fss.rml}
   r_M(\lambda,\sigma;L) := 1 - \frac{ \chi_2^{(M)}
   (\widetilde\kappa,\lambda,\sigma;L) }
   { \chi_2^{(M)}(\widetilde\kappa,\lambda,\sigma;\infty) },
\ee
we know that
$r_\infty(\cdot;L)>0$ for 2nd order and
$r_\infty(\cdot;L)<0$ for 1st order transitions, if $L$ lies
within the bounds as explained above. The convergence of the series
\eqn{fss.chi2} as
$M\to\infty$ ensures the same behaviour for finite, but sufficiently
large $M$.

\begin{figure}[htb]
\caption{\label{monotony} Dependence of the ratios
$r_{20}(\lambda,\sigma;L)$ as defined by \eqn{fss.rml}
on $L=(L_1L_2L_3)^{1/3}$, at the example of the
$\mathbb{Z}_2$-model. Two points of the phase space have been
chosen.
\hfill\break
1. ($+$) $\lambda=15/1.1$ and $\sigma=15$.
The transition is 1st order, $r_{20}<0$.
\hfill\break
2. ($\square$) $\lambda=3$ and $\sigma=1$.
The transition is 2nd order, $r_{20}>0$.
}

\setlength{\unitlength}{0.8cm}
\begin{picture}(12.0,12.0)

\epsfig{file=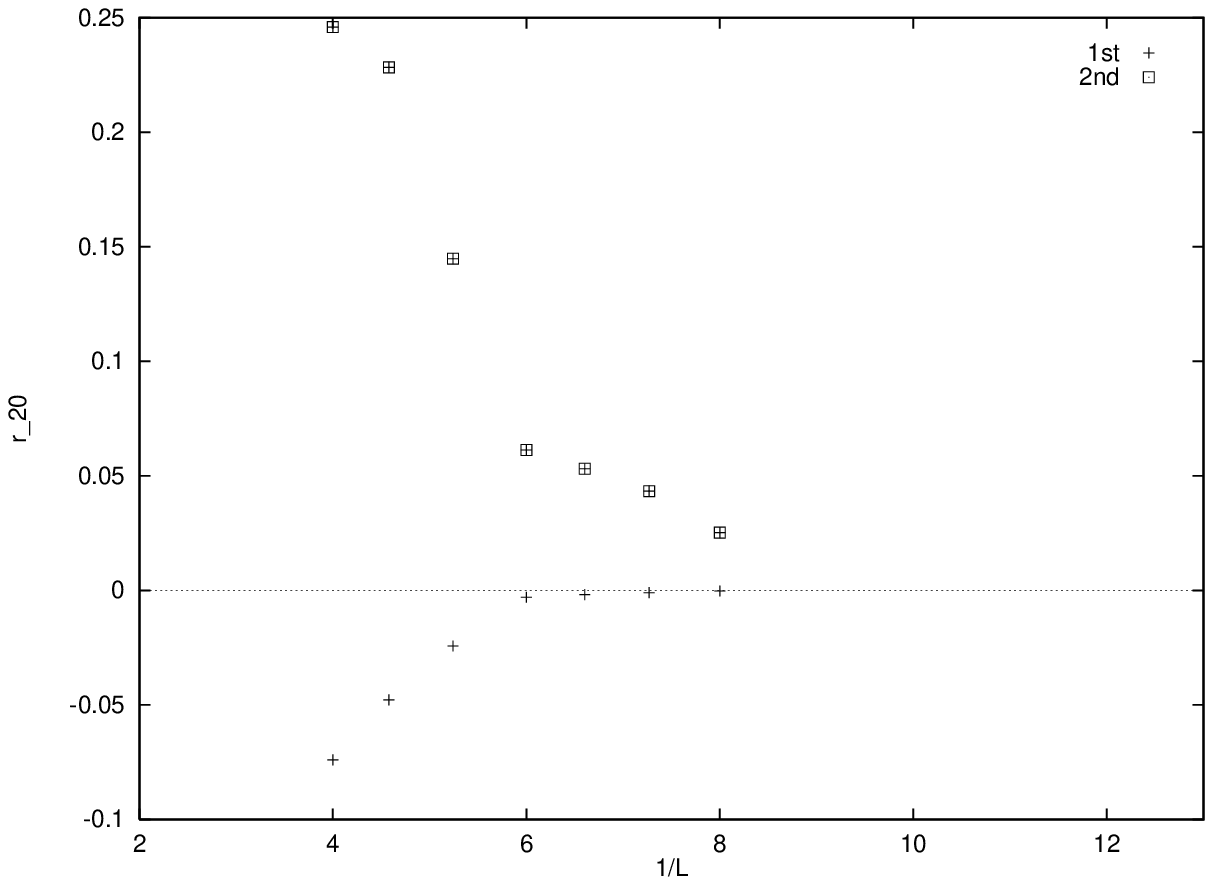}

\end{picture}

\end{figure}


Fig.~\ref{monotony} shows an application of the monotony criteria
to the 1-component model.
We have plotted the volume dependence of the ratios
$r_M(\lambda,\sigma;L)$ for a truncation at order $M=20$ and
for various lattice
sizes $L=(L_1L_2L_3)^{1/3}$, at two points of the bare coupling
constant space. One point is well inside  the 1st order region,
the other one well inside the 2nd order part of the phase
diagram, the different areas have been identified by the infinite
volume series as discussed
in the last subsection.
Clearly the sign of $r_{20}(\cdot;L)$ is different in
both regions of phase space.
It is positive for the 2nd order transition and
negative for the 1st order transition.
The approach to the infinite volume limit where
$r_M(\cdot;\infty)=0$ is fast.

Next we want to demonstrate how one can utilize the finite volume criteria
with HPE to get a better localization of the tricritical region.
Let us consider the O(4) model and determine
the behaviour of the ratios $r_M$ along the line $\lambda = (1/2)\sigma$.
{}From the infinite volume analysis of the last subsection
we had obtained
$\sigma_t\simeq 9.0$ as an estimate for the tricritical coupling $\sigma_t$.

\begin{figure}[htb]
\caption{\label{ratios} The ratios
$r_M(\lambda,\sigma ;L)$ as function of $\sigma$ for $\lambda = \sigma/2$,
and $L=4$, i.e. on a $4\times 4\times 4$-lattice.
}

\setlength{\unitlength}{0.8cm}
\begin{picture}(12.0,12.0)

\epsfig{file=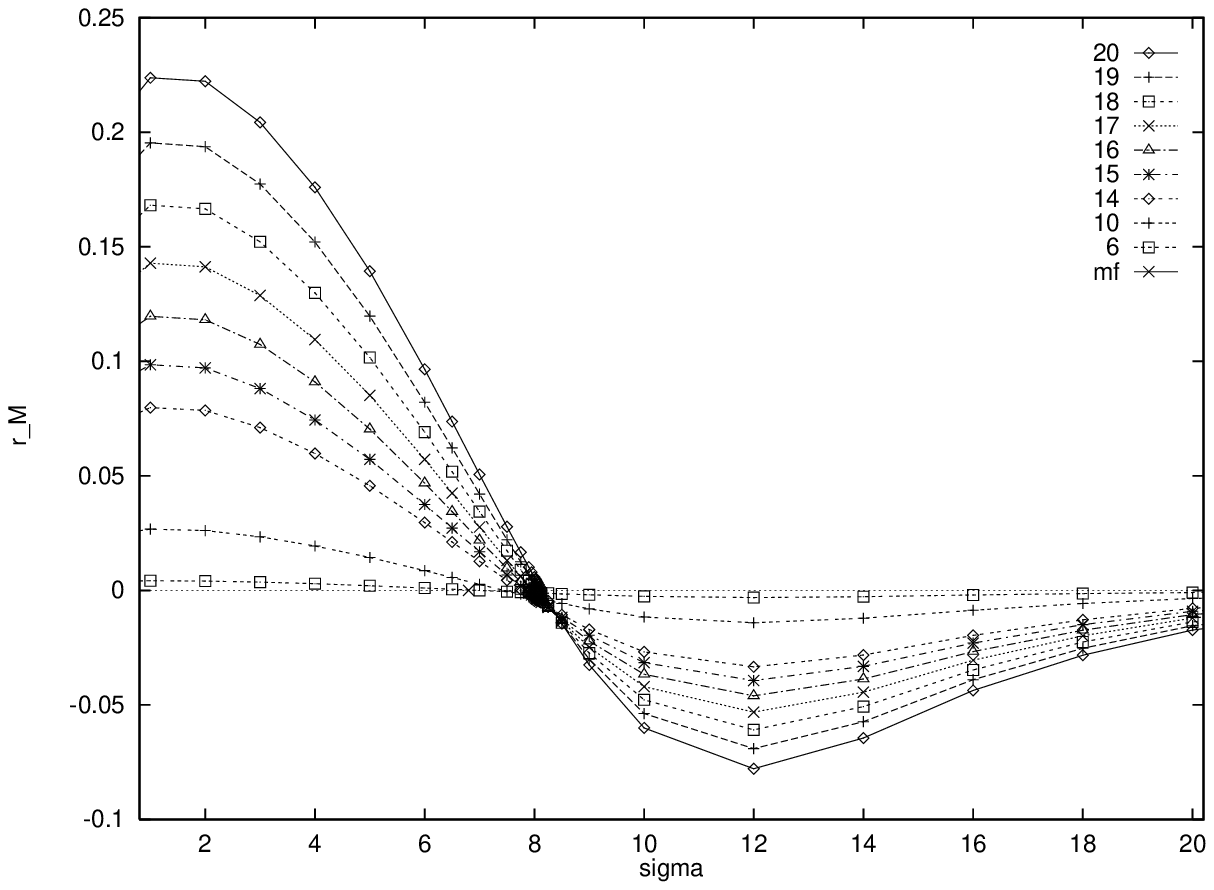}

\end{picture}

\end{figure}


Since $r_M(\lambda,\sigma ;L)>0$ for 2nd order and $<0$ for 1st order
transitions, the tricritical point should be localized at the zero of
$r_M(\lambda,\sigma ;L)$
between these two ranges (with nonvanishing slope, i.e.
$\partial r_M(\sigma/2,\sigma ;L)/\partial\sigma\not=0$),
suitably extrapolated to $M,L \to\infty$.

\begin{figure}[htb]
\caption{\label{mon19} Dependence of the ratios
$r_{19}(\sigma /2,\sigma;L)$ as defined by \eqn{fss.rml}
on $\sigma$ for various volumes.
}

\setlength{\unitlength}{0.8cm}
\begin{picture}(12.0,12.0)

\epsfig{file=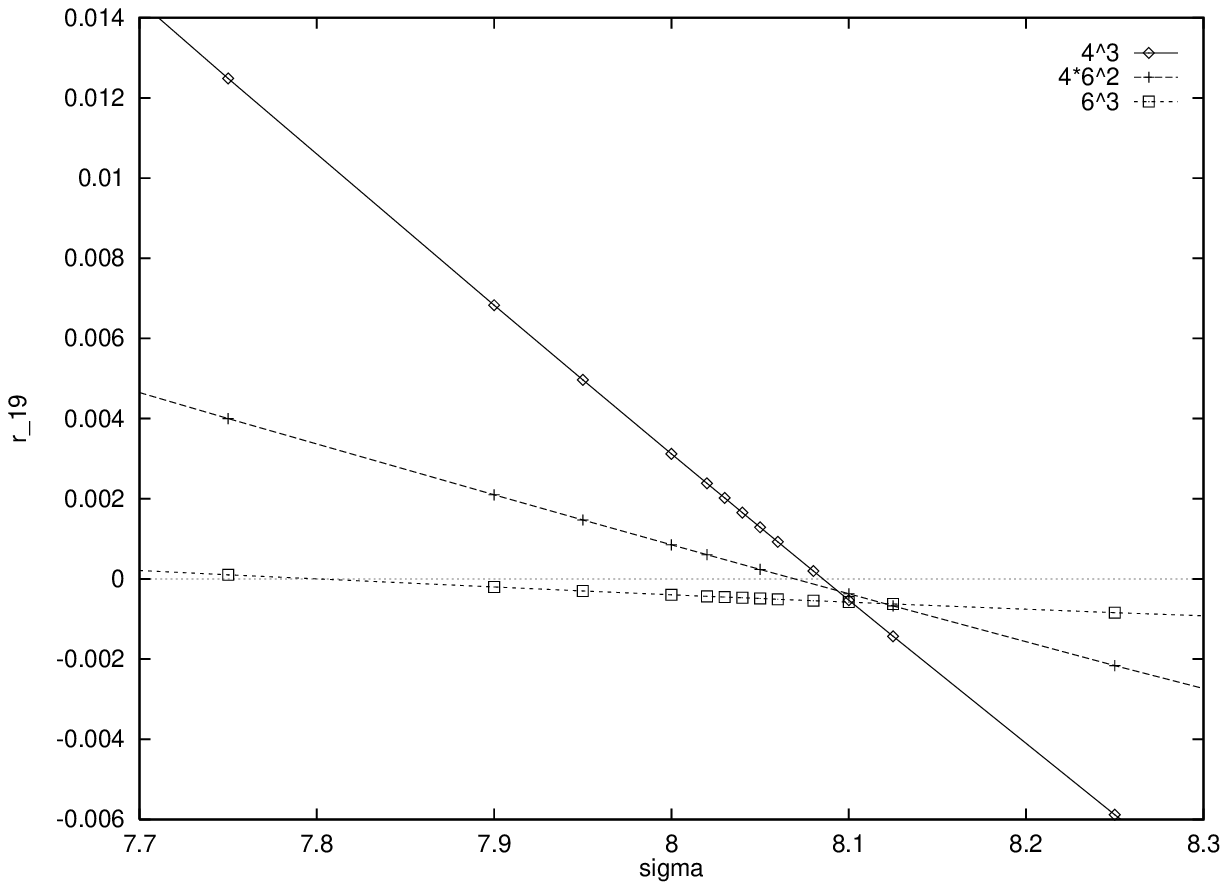}

\end{picture}

\end{figure}

Fig.~\ref{ratios} shows the ratios
$r_M(\sigma / 2,\sigma ;L)$ as function of $\sigma$ for $L=4$ and various $M$
between $0$ (corresponding to the mean field approximation) and $20$.
The intersections of the curves with the $r=0$-axis lie in the range of
$8\leq\sigma\leq 8.5$.
The zeroes $\sigma_M(L)$, defined by
$r_M(\sigma_M(L) / 2,\sigma_M(L) ;L)=0$, depend on the order $M$
at which the susceptibility series of $\chi_2$ have been truncated
and on the lattice size $L$. The dependence on $L$ for fixed $M$
of $r_M(\lambda,\sigma ;L)$ is shown in Fig.~\ref{mon19}.

Thus a final localization of the tricritical coupling $\sigma_t$ needs
an extrapolation in $L$ and $M$ to infinity.
Clearly both extrapolations are not independent of each other.
We should expect that a comparision of two ratios is sensible, i.e.
\be \label{exa.corr}
  r_M(\cdot ; L) \simeq r_{M^\prime}(\cdot ; L^\prime )
\ee
if lattice sizes $L,L^\prime$ and truncation $M,M^\prime$ satisfy
\be \label{exa.ratio}
  \frac{M}{L} = \frac{M^\prime}{L^\prime}.
\ee
The reason is that $M/L$ is the maximal number of times a graph
contributing to the series of $\chi_2^{(M)}(\cdot;L)$
can wind around the volume. Eq.~\eqn{exa.ratio} then ensures that the
remaining $L$-dependence becomes independent of $M$ for sufficiently
large $M$.

\begin{figure}[htb]
\caption{\label{reg1} The solution $\sigma_M(L)$ of
the equation $r_M(\sigma_M(L)/2,\sigma_M(L);L)=0$ for $L=4$,
plotted against $1/M$,
$M$ is the order of truncation of the suceptibility series.
Regression is shown as a linear function of $1/M$ for $M\geq M_{min}$
with $M_{min}=14,16,18$.
}

\setlength{\unitlength}{0.8cm}
\begin{picture}(12.0,12.0)

\epsfig{file=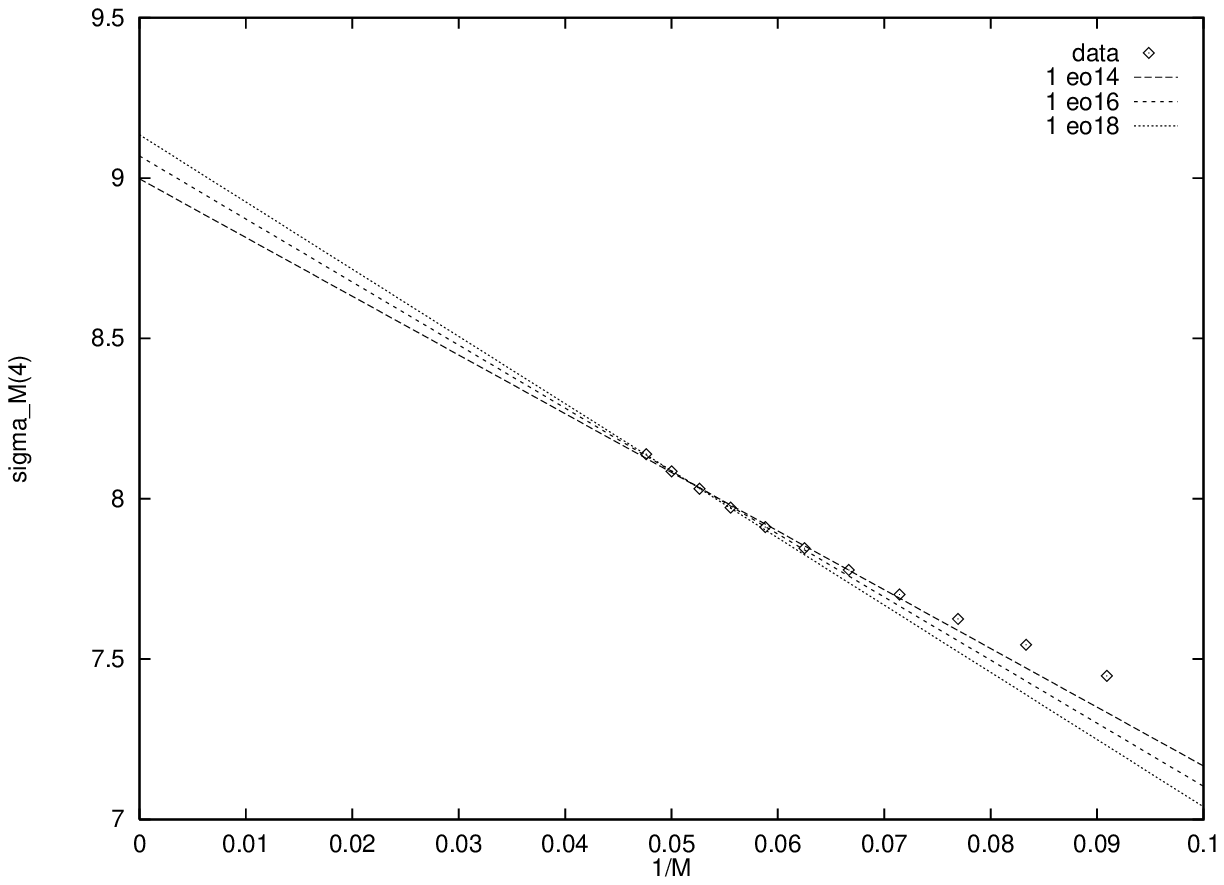}

\end{picture}

\end{figure}


\begin{figure}[htb]
\caption{\label{reg2} The "scaled" solutions
$\sigma_t(M_{min}L/4,L)$, plotted against
$1/M_{min}$, for $L=4$ and $L=6$.
Within the error bars, the data are on a straight
line, thus confirming the
assumptions made in connection
with Eqns.~\eqn{exa.corr} and \eqn{exa.ratio}.
Linear regression in $1/M_{min}$ gives
the tricritical point.
}

\setlength{\unitlength}{0.8cm}
\begin{picture}(12.0,12.0)

\epsfig{file=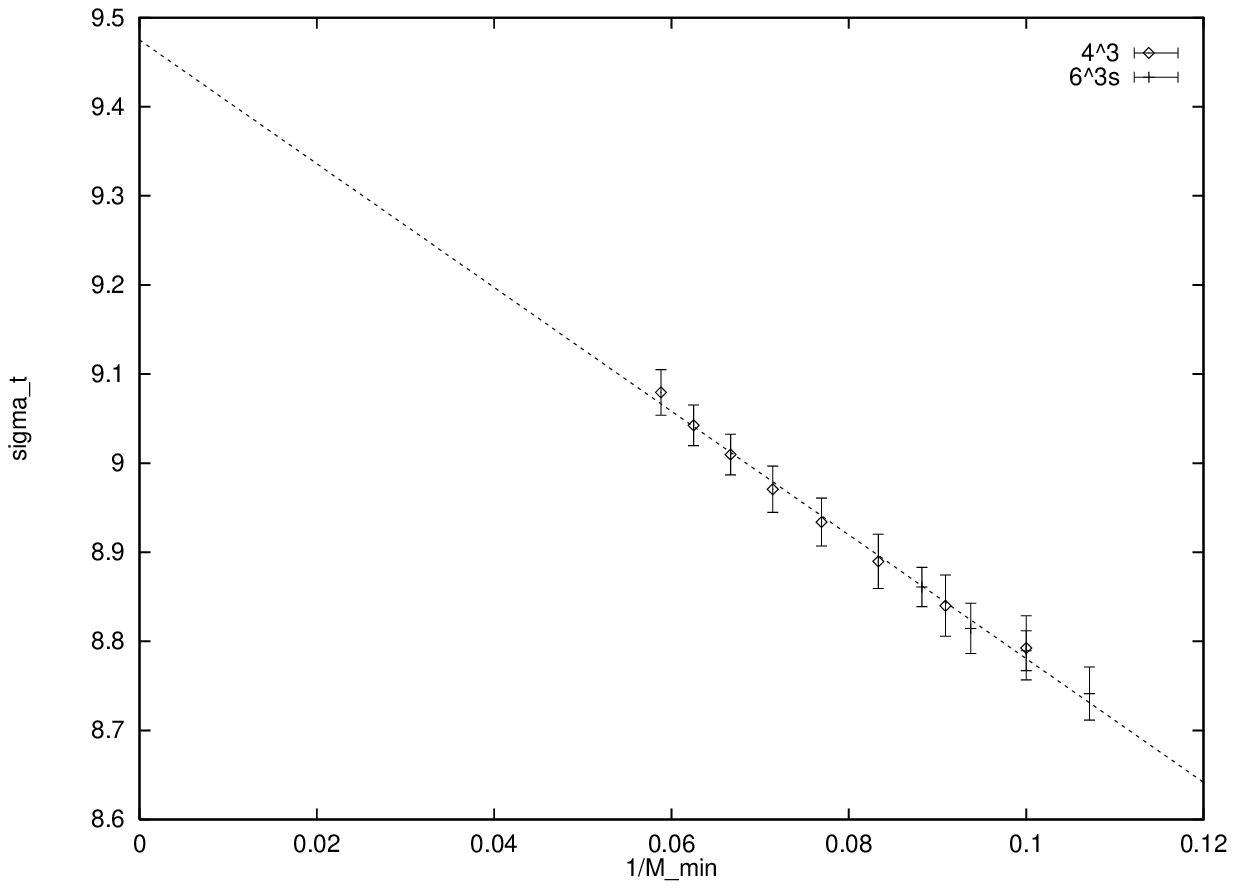}

\end{picture}

\end{figure}


Fig.~\ref{reg1} shows the data $\sigma_M(L)$ obtained on a
$4^3$-lattice ($L=4$), and for $11\leq M\leq 20$,
as a function of $1/M$.
The curves show the regression
\[
   \sigma_M(4) = \sigma_t(M_{min}, L=4) +
   \frac{\delta(M_{min})}{M},
\]
as obtained for $M_{min}\leq M \leq 20$, and for
$M_{min}=14,16,18$.
In Fig.~\ref{reg2} we show the second regression for the resulting
$\sigma_t(M_{min} L/4, L)$ in $1/M_{min}$,
for $L=4,6$
\[
  \sigma_t( M_{min} \frac{L}{4} , L ) = \sigma_t
  + \frac{\epsilon}{M_{min}},
\]
leading to the final
prediction of $\sigma_t$.
We have scaled the adjusted $M_{min}$ according to
\eqn{exa.ratio}.
Note that both data sets, on the $4^3$ and
on the $6^3$-lattice, fall on the same
straight line within the error bars.
We obtain
\be \label{tric}
   \sigma_t = 9.454(49) .
\ee

In passing we remark that the accuracy of $\sigma_t$ has been increased
by at least one order of magnitude compared to the infinite volume
analysis (Fig.~\ref{nueta}).
It should be noticed that the volume independence of $\chi_2$ in the
scaling region of the tricritical point just confirms the validity of
a mean field analysis of tricritical exponents. A mean field analysis
is volume independent by construction. The expected
volume independence of $\chi_2$
is confirmed within the error bars.

In the remainder of this section we discuss
the volume dependence of the radius of convergence
and of the effective potential.

{\it Shift and scaling of $\kappa_c(\lambda,\sigma ;L)$.}
We define $\kappa_c(\lambda,\sigma ;L)$ in the finite and
infinite volume as the radius of convergence of
the susceptibility series, in particular of $\chi_2$,
which is known including the 20th order.
Some data for $\kappa_c(L)$ are listed in
Table \ref{exa.kappal} for a three-dimensional O(1) and O(4) model. The
couplings $(\lambda,\sigma)$ for each N ($N=1,4$) have been chosen
deeply in the 1st order transition region.

{}From a finite size scaling analysis one expects (cf.~Sect.~\ref{mon.title})
that the data should fit with a regression in $L$ according to
\[
   | \kappa_c(\cdot;L) - \kappa_c(\cdot;\infty) | \simeq
   \ln{c} - y_T \ln{L}
\]
for large $L$,
with some constant $c$ and a critical exponent $y_T$. For the
$\mathbb{Z}_2$ model we obtain in the 1st order transition region
($\lambda=15/1.1, \sigma=15.0)$ according to
Table \ref{exa.kappal}
\bea
   \ln{c} &=& 4.57(57) \nonumber \\
   \label{exa.yt2}
   y_T &=& 6.21(32) \nonumber
\eea
with $\chi^2/df$=0.025. Thus the scaling behaviour is consistent with
$y_T=2D=6$.
It confirms the behaviour which has been predicted for
the shift of the critical coupling determined as the maximum of the
susceptibility in a class of models
which cover the $\mathbb{Z}_2$ model
\cite{borgs2}. Note that it is in disagreement with the
 Gaussian two-peak model \cite{binder}, predicting a
leading finite size correction proportional to $L^{-D}$, which one might
have expected.


\begin{table}[htb]
\caption{\label{exa.kappal} Radius of convergence
$\kappa_c(L)$
of the HPE series for the 1st order region.
It is determined from the
2-point suceptibility series, as
described in Section 4.2.
Data are given for the 3d $\mathbb{Z}_2$ and O(4) model,
for various volumes $L^3$.
}
\vspace{0.5cm}

\begin{center}
\begin{tabular}{|r|c||c|}
\hline \hline
$ \; $ & $\mathbb{Z}_2$   &   $O(4)$   \\ [0.5ex] \cline{2-3}
$ \: $ & $\lambda = 13.64$ , $\sigma = 15.0$ &
$\lambda = 6.0$ , $\sigma = 12.0$ \\ [0.5ex] \cline{2-3}
$ L $ & $\kappa_c(L)$ &
$\kappa_c(L)$   \\
[0.5ex] \hline
$\infty$ &  0.51047(1)&   0.84462(10) \\ [0.5ex] \hline
  4      &  0.49381   &   0.79531  \\
  6      &  0.50886   &   0.83851  \\
  8      &  0.51025   &   0.84416  \\
  10     &  0.51047   &   0.84436  \\
  12     &  0.51047   &   0.84447  \\ \hline
\end{tabular}

\end{center}

\end{table}

The same scaling behaviour is found for the O(4) model in
the 1st order transition region. Here we get
(Table \ref{exa.kappal})
\bea
   \ln{c} &=& 4.57(120) \nonumber \\
   y_T &=& 5.55(59)  \nonumber.
\eea
Thus a leading correction proportional to
$L^{-3}$ lies clearly outside the error bars. This result is remarkable,
because the $O(4)$-model (which is a Heisenberg model in the large
coupling limit) is not covered by the analysis of Borgs, Imbrie, Kotecky
and Miracle-Sole
\cite{borgs,borgs2}. Hence the vanishing of the coefficient of the linear
term in a large volume expansion of the susceptibilities
in powers of $1/L^3$ seems to be a
universal feature of a large class of models.


\begin{table}[htb]
\caption{\label{exa.kappal2} Radius of convergence
$\kappa_c(L)$
of the HPE series for the 2nd order region in
the $\mathbb{Z}_2$ model
for various volumes $L^3$.
}
\vspace{0.5cm}

\begin{center}
\begin{tabular}{|r|c|}
\hline \hline
$ \; $ & $\lambda = 3.0$ , $\sigma = 1.0$  \\
$ L $ & $\kappa_c(L)$    \\
[0.5ex] \hline
$\infty$ &  0.17316(4) \\ [0.5ex] \hline
  6      &  0.17393   \\
  8      &  0.17339   \\
  10     &  0.17327   \\
  12     &  0.17327   \\ \hline
\end{tabular}

\end{center}

\end{table}

A measurement of the scaling behaviour of $\kappa_c(L)$ in the
2nd order region of the O(4) model was not conclusive, because the
shift of $\kappa_c$ occurs in the 4th or 5th digits, hence the
finite size effect is hidden in the error. In the 2nd order region
of the $\mathbb{Z}_2$ model we measure a scaling,
which is best fitted by an ansatz (cf.~Table \ref{exa.kappal2})
\[
   \kappa_c(\cdot ;L) = \kappa_c(\cdot ;\infty) +
   \frac{c}{L^{y_T}}
\]
with $y_T = 2/\nu$, and
\bea
   \ln{c} &=& -2.25(129) \nonumber \\
   y_T &=& 2.83(59)  \nonumber
\eea
with $\chi_2/df$=0.1.
{}From a renormalization group analysis one expects a leading
scaling correction proportional to $L^{-1/\nu}$. Also here the finite
size effect is not large compared to the error in $\kappa_c$,
thus the disagreement with
an $L^{-1/\nu}$-behaviour may be due to the errors in $\kappa_c$ and
the uncertainty in the involved extrapolations.

We conclude the discussion of $\kappa_c(L)$ with a conjecture concerning
the monotony behaviour.

{\it Monotony in $\kappa_c(L)$.}
The results of Table \ref{exa.kappal} for $\kappa_c(L)$ exhibit a further
characteristic distinction between 1st and 2nd order transitions.
Since we are not aware of a general proof, we leave it as a

{\bf Conjecture}: The convergence radius $\kappa_c(L)$ of the series
expansions is monotonically decreasing with $L$ for 2nd order
transitions, and monotonically increasing for 1st order transitions.

\vskip 1cm

{\it The effective potential as function of $L$.}
Similarly we have measured $V_{eff}$, Eq.~\eqn{mon.effpot},
for two points well
inside the supposed 1st and 2nd order transition regions of the O(4) model
($\lambda=6.0$, $\sigma=12.0$, and $\lambda=0.\overline{90}$, $\sigma=1.0$,
respectively). The $\chi_{2n}^{1PI}$ entering Eq.~\eqn{mon.effpot}
have been evaluated to 16th order in $\kappa$
for $n=1,2,3$ in a finite and infinite volume.
For the 2nd order point $\lambda=0.\overline{90}$, $\sigma=1.0$,
and for volumes $L^3$ with $L=4,6$ and infinity, $V_{eff}$ is
convex in the symmetric phase up to a resolution of $10^{-8}$.

\begin{figure}[htb]
\caption{\label{Veff} Volume dependence of the effective
 potential $V_{\rm eff}(\Phi)$
 on lattices with varying size $L^3$.
 The barrier height decreases with $L$. The curves are obtained for
 $L=4,6,\infty$. The parameters are $\lambda=6.0$,
 {$\sigma=12.0$} in the 3d O(4)-model.
}

\setlength{\unitlength}{0.8cm}
\begin{picture}(12.0,12.0)

\epsfig{file=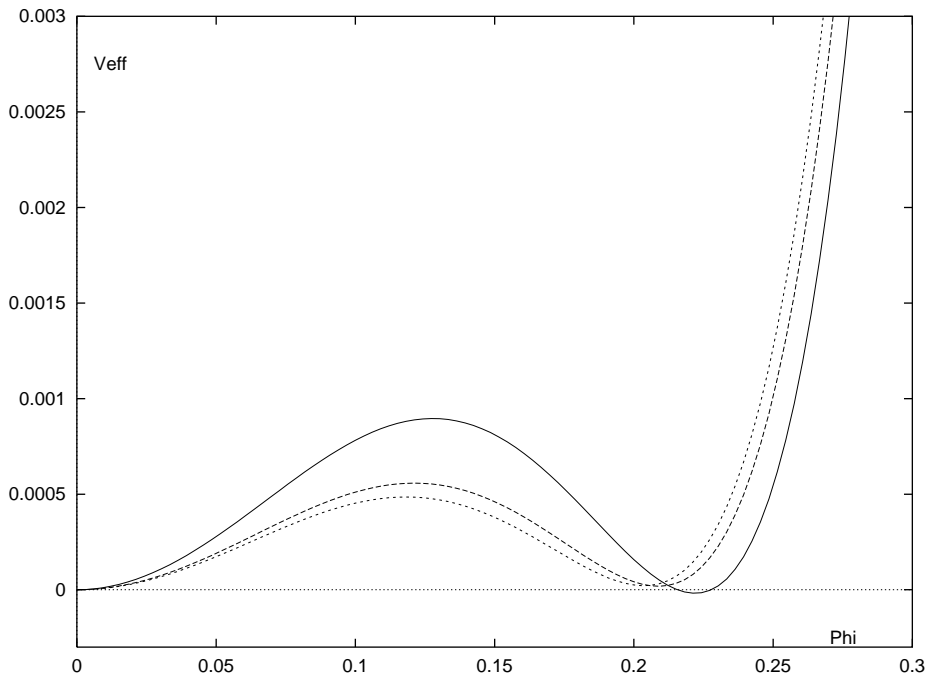}

\end{picture}

\end{figure}


In the first
order case (Fig.~\ref{Veff}) $V_{eff}$ is nonconvex in the symmetric phase
with a barrier height decreasing with increasing $L$. The three curves
correspond to a $4^3$,  $6^3$ and $\infty^3$ lattice, respectively.
Note that the nonconvex shape which
is even seen for the $\infty^3$ lattice (lowest barrier in Fig.~\ref{Veff})
must be attributed to the approximation scheme,
i.e. to the truncation at order $M=16$. (As mentioned above, the truncation
of the series expansion acts similarly to a finite volume cut-off. The
finite volume leads to a nonconvex shape only in the 1st order case.
Thus it is not surprising that we find a convex shape in the symmetric
phase for all volumes in the 2nd order case, in spite of the same
truncation in $M$.)

The coexistence of minima leading to the same value of $V_{eff}$
defines a critical coupling $\widetilde\kappa_c(L)$ which need not
agree with the finite
volume convergence radius $\kappa_c(L)$, unless the truncated expansions of
$V_{eff}(\Phi)$ in $\Phi$ and of $\chi_{2n}^{1PI}$ in $\kappa$ are
extrapolated to infinity. For parameters chosen as in Fig.~\ref{Veff} we find
$\widetilde\kappa_c(4)=0.8311$, $\widetilde\kappa_c(6)=0.8475$,
$\widetilde\kappa_c(\infty)=0.8488$, in contrast to
$\kappa_c(4)=0.79531$, $\kappa_c(6)=0.83851$,
$\kappa_c(\infty)=0.84462$, cf. Table~\ref{exa.kappal},
determined as the radius of convergence of the series
expansions, extrapolated to infinite $M$.

{\it Tricritical parameters from $V_{eff}$.}
Vanishing coefficients of the $\Phi^2$ and $\Phi^4$-terms associated with a
qualitative change in the shape of $V_{eff}$ provide a further
possibility for localizing the tricritical couplings. The analytical
dependence of the coefficients on $\lambda$ and $\sigma$ is rather indirect.
It is easier to determine $\lambda_{t}$, $\sigma_{t}$ by the first
occurrence of a nonconvex shape of $V_{eff}$ in the symmetric phase,
coming from the 2nd order transition region. The highest order, which
is so far available for the six-point susceptibility $\chi_6^{1PI}$, entering
the $\Phi^6$-coefficient of $V_{eff}$, is $16$. The number of
contributing graphs to order $16$ is comparable
to the number of graphs contributing to  the two-point susceptibility
to order 20 and the four-point susceptibility to order 18.
An inclusion of higher powers in $\Phi$,
say $\Phi^8, \Phi^{10}$-terms, would further reduce the order in
$\kappa$ which is tractable. Thus we retain
from further extrapolations, but give  bounds on $\sigma_{t}$, derived
from $V_{eff}(\Phi)$ to $O(\kappa^{16})$. They are
\be \label{boundveff}
   9.75 \leq \sigma_t \leq 10.0 .
\ee
The resolution in $V_{eff}$, within
which no nonconvex shape was seen up to $\sigma=9.75$, was $10^{-7}$.
Compared to the more reliable result of Eq.~\eqn{tric},
the evaluation of $V_{eff}$
seems to lead to an upper bound on $\sigma_{t}$, given
by Eq.~\eqn{boundveff}.

%
%
%

\section{Summary and Outlook}

In this paper we have generalized hopping parameter expansions from
an infinite to a finite volume. The combination of performing the
expansions in a finite volume and to a high (20th) order in the
expansion parameter has turned out as a useful computational technique
to approach the critical region from the symmetric phase and, in
addition, to characterize the type of transition. First and second
order transitions have been distinguished by various criteria:
the monotony criterion refering to the $L$-dependence of
response functions (here illustrated at order parameter
susceptibilities),the scaling and monotony of the radius of
convergence $\kappa_c(L)$ as function of the linear lattice size
$L$, and the effective potential as function
of $L$. In particular, it is the different L-dependence of
the order parameter susceptibility for 1st and 2nd order transitions
in the scaling region
(close to but not at the transition point)
which allows us to localize tricritical
points. The monotony criterion can be applied to Monte Carlo
simulations as well, the involved two volumes should be
sufficiently large, but both may be
finite.

We have applied these methods to renormalizable O(N) models in three
dimensions.
The plateau structure in the critical exponent $\nu \eta$ in the
infinite volume has revealed two universality classes belonging to
an $O(N)$-Heisenberg model and to a Gaussian model.
Apart from the "trivial" Gaussian behaviour at vanishing four and
sixpoint couplings,
we get Gaussian exponents along a tricritical line separating
1st and 2nd order domains.
The existence of the 1st order domain and the tricritical line
is based on the presence of the $\Phi^6$ self-interaction.
The O(N) symmetry alone does not determine the critical behaviour.

Several extensions are at hand. The first one is from 3 dimensions to
field theories in four dimensions at finite temperature.
For a $\Phi^{4}+\Phi^{6}$-type theory in four dimensions at finite
temperature  we expect qualitatively the same infrared behaviour
and phase structure as for
three dimensions, but different values for the critical couplings.
A check of the
supposed dimensional reduction from four to three dimensions
is of particular interest in connection with the electroweak phase
transition.
$\Phi^{6}$ terms in the dimensionally reduced $SU(2)$-Higgs
model are usually argued to be irrelevant even at the
transition to the spontaneously broken phase,
and hence dropped \cite{kajantie}. In
our extension of the previous investigations we will keep the $\Phi^6$
term in a 4d effective scalar theory at finite temperature,
which is derived from the
underlying $SU(2)$-Higgs model by integrating out the gauge
field degrees
of freedom. The phase structure of the effective
scalar theory will then be studied
in a finite and infinite volume.
Hopping parameter expansions are supposed
to work the better the smaller $\kappa$, thus
the larger the Higgs masses.
Hence this investigation complements the range of Higgs masses
which has been available in recent Monte Carlo simulations \cite{fodor}.
One of our aims is to find
the {\it critical} Higgs mass above which the electroweak phase transition
ceases to exist. (In case that the {\it physical} Higgs mass lies above the
critical Higgs mass, it is bad news for an explanation of the observed
baryon number asymmetry in the universe. The necessary ingredient for
an out-of-equilibrium situation can no longer be provided by the
electroweak transition, if the "transition" turns out to be truly a
smooth crossover phenomenon.)

A further application of our computational tools are (tri)critical
phenomena in statistical physics. The order of the transition in a
superconductor of type II has been recently under debate
(\cite{ra} and references therein.)
The existence of a tricritical point for a suitable Ginzburg-Landau
parameter has been conjectured \cite{bar}, but a proof of its existence is
still outstanding.
Work in both directions is in progress.


\begin{appendix}

%
%
\section{\label{app1.0}Large coupling limit of O(N) lattice models}

For $N\geq 1$,
on a $D$-dimensional hypercubic lattice $\Lambda$, we consider
the partition function
\[
  Z = \int \cD\Phi \exp{(-S(\Phi))}
\]
with corresponding expectation values
\be \label{app1expp}
  < P >_{\lambda,\sigma} =
   \frac{1}{Z} \int \cD\Phi P(\Phi) \exp{(-S(\Phi))},
\ee
where
\[
   \cD\Phi = \prod_{x\in\Lambda} d^N\Phi_x
\]
and
\bea
    S(\Phi) & = & -\frac{1}{2} \sum_{x\not=y} v_{xy} \Phi_x \cdot\Phi_y
     + \sum_x \stackrel{\circ}{S}(\Phi_x) , \nonumber\\
      \stackrel{\circ}{S}(\Phi) & = & \Phi^2 + \lambda (\Phi^2-1)^2
       + \sigma (\Phi^2-1)^3.\nonumber
\eea
Measure and action are globally O(N) invariant. The observable
$P$ should be appropriately bounded so that the integrals exist.
Fields at different lattice sites $x$ and $y$ interact by the
hopping coupling  $v_{xy}$, which is assumed to obey the
following conditions
\bea \label{app1vp}
  v_{xy} = v_{yx} & = & v(x-y) , \nonumber \\
  v(0) & = & 0 , \nonumber \\
  \sum_{y\in\Lambda} v_{xy} & < & \infty , \\
  \sum_{x,y\in\Lambda} v_{xy} (\Phi_x-\Phi_y)^2 & \geq & 0.\nonumber
\eea
As an example, these conditions are satisfied
for the pure nearest neighbour interaction,
where
\be \label{app1.nn}
   v_{xy} = 2\kappa \sum_{\mu=0}^{D-1}
   ( \delta_{x,y+\widehat\mu} + \delta_{x,y-\widehat\mu} ),
\ee
with $\widehat\mu$ denoting the unit vector in $\mu$-direction.
We consider \eqn{app1expp} in the limit $\sigma\to\infty$
with $\lambda = \alpha \sigma$ and $\alpha$ a fixed real number,
i.e.
\[
  < P >_\alpha = \lim_{\sigma\to\infty} < P >_{\alpha\sigma,\sigma}
\]
for
\be \label{app1.actal}
  \stackrel{\circ}{S}(\Phi) = \Phi^2 + \sigma (\Phi^2-1)^2
        ( \Phi^2-1+\alpha ).
\ee
The hopping parameters $v_{xy}$ are considered as fixed.
We have to select all
field configurations that minimize the action in this limit.
It is convenient to parametrize the fields according to
\[
    \Phi_x  =  {u}_x v_x, \;
    {u}_x\in S_{N-1} \; \mbox{(the N-1 sphere)}, v_x\geq 0.
\]

\begin{lemma}\label{app1l1}
As $\sigma\to\infty$ we get the following behaviour of
expectation values $P$ in dependence on $\alpha$
\begin{description}
\item{$\underline{\alpha > 1}$:}
\[
  < P >_\alpha = \cN_\alpha \prod_{x\in\Lambda} \left( \int_{S_{N-1}}
   d\Omega_{N-1}(u_x) \right) \; P(u) \exp{(-\widetilde S(u))},
\]
\[
   \widetilde S(u) = -\frac{1}{2} \sum_{x,y\in\Lambda}
   v_{xy} u_x\cdot u_y .
\]
\item{$\underline{\alpha = 1}$:}
\[
  < P >_1 = \cN_1 \prod_{x\in\Lambda}
   \left( \sum_{v_x=0,1} ( \int_{S_{N-1}} d\Omega_{N-1}(u_x) \delta_{v_x,1}
    + \delta_{v_x,0} ) \right)
   P(uv) \exp{(-\widetilde S(u,v))},
\]
\[
   \widetilde S(u,v) = \sum_{x\in\Lambda} v_x^2
    -\frac{1}{2} \sum_{x,y\in\Lambda}
   v_{xy} v_x v_y u_x\cdot u_y .
\]
\item{$\underline{-3 < \alpha < 1}$:}
\[
  < P >_\alpha = P(0) .
\]
\item{$\underline{\alpha = -3}$:}
\bea \nonumber
  < P >_{-3} = \cN_{-3} \prod_{x\in\Lambda}
   && \left( \sum_{v_x=0,1}
    ( \int_{S_{N-1}} d\Omega_{N-1}(u_x) \delta_{v_x,1}
    + \delta_{v_x,0} ) \right)  \\ && \nonumber \\
   \nonumber
   && \qquad P(\sqrt{3}uv) \exp{(-\widetilde S(u,v))},
\eea
\[
   \widetilde S(u,v) = \sum_{x\in\Lambda} 3 v_x^2
    -\frac{3}{2} \sum_{x,y\in\Lambda}
   v_{xy} v_x v_y u_x\cdot u_y .
\]
\item{$\underline{\alpha < -3}$:}
\[
  < P >_\alpha = \cN_\alpha \prod_{x\in\Lambda} \left(\int_{S_{N-1}}
   d\Omega_{N-1}(u_x) \right) \;
   P((1-\frac{2\alpha}{3}) u) \exp{(-\widetilde S(u))},
\]
\[
   \widetilde S(u) = -\frac{1}{2} \sum_{x,y\in\Lambda}
   (1-\frac{2\alpha}{3}) v_{xy} u_x\cdot u_y .
\]

\end{description}
The $\cN_\alpha$ are positive normalization factors independent of $P$
such that $<1>_\alpha=1$,
$d\Omega_{N-1}$ is the standard measure on the sphere.
\end{lemma}

For $\alpha>1$ and $\alpha<-3$ we obtain the O(N) Heisenberg model
 (Ising model for N=1).
For $-3<\alpha <1$ the lattice model becomes
completely decoupled. At the boundary points
$\alpha=1$ and $\alpha=-3$ the result are
"diluted" $O(N)$ models, i.e. $O(N)$ models
with additional occupation number variables $v_x \in \{0,1\}$.

\underline{Outline of the proof:}
The properties \eqn{app1vp} ensure that the minimizing field
configurations of the action $S(\Phi)$ are translation invariant.
Thus it is sufficient to determine the minima of
\[
    F(\Phi) = (\Phi^2-1)^2 (\Phi^2-1+\alpha),
   \; \Phi\in{\bf R}^N
\]
for the various values of $\alpha$. Finally the saddle point expansion
for the different
cases yields the lemma.

An alternative way to study the large coupling limit is to
perform it termwise in the HPE series of correlation functions.
We specialize to the nearest neighbour interaction \eqn{app1.nn}.
The only way the coupling constants $\lambda$ and $\sigma$ enter
the linked cluster expansion
is via the connected one-point vertex couplings
$\stackrel{\circ}{v}_{2n}^c(\lambda,\sigma)$,
defined in Eqn.~\eqn{lce.14}. The connected one-point vertex couplings
are
related to the full one-point couplings $\stackrel{\circ}{v}_{2n}$,
defined by
\be
     \stackrel{\circ}{v}_{2n} = \frac{ \int d^N\Phi
       \Phi_1^{2n} \exp{(-\stackrel{\circ}{S}(\Phi))} }
       { \int d^N\Phi
        \exp{(-\stackrel{\circ}{S}(\Phi))} } ,
\ee
by the identity \eqn{lce.vcc}.
At any finite order $l$, the coefficient of $\kappa^l$
is a polynomial in the $\stackrel{\circ}{v}_{2n}^c$,
hence in the $\stackrel{\circ}{v}_{2n}$.
Invoking a saddle point expansion again, we obtain

\begin{lemma}\label{app1l2}
Let $\lambda = \alpha\sigma$ with
$\sigma>0$, $-\infty <\alpha <\infty$.
Define for nonnegative integers $N,k$
\[
   \cA_{N,k} \; = \; (2k-1)!! \; \frac{\Gamma(\frac{N}{2})}
    {\Gamma(\frac{N}{2}+k) 2^k}.
\]
As $\sigma\to\infty$ we get for every $k>0$
\begin{description}
\item{$\underline{\alpha > 1}$:}
\[
   \stackrel{\circ}{v}_{2k}\; = \;
   \cA_{N,k} \; + \; O(\sigma^{-\frac{1}{2}}).
\]
\item{$\underline{\alpha = 1}$:}
\[
  \stackrel{\circ}{v}_{2k}\; = \; \cA_{N,k} \cdot
    \; \left\{
   \begin{array}{r@{\qquad ,\quad} l }
    \frac{1}{e + 1} + O(\sigma^{-\frac{1}{2}}),
       & {\rm for \; N=1,} \\
    1 + O(\sigma^{-\frac{1}{2}})& {\rm for \; N\geq 2.}
   \end{array} \right.
\]
\item{$\underline{-3 < \alpha < 1}$:}
\[
   \stackrel{\circ}{v}_{2k}\; = \;
   \frac{(2k-1)!!}{2^k} \;
   \frac{1}{\left( (3-2\alpha)\sigma \right)^k}
    \; ( 1 + \; O(\sigma^{-\frac{1}{2}}) ).
\]

\item{$\underline{\alpha = -3}$:}
\[
  \stackrel{\circ}{v}_{2k}\; = \; \cA_{N,k} \cdot
    \; \left\{
   \begin{array}{r@{\qquad ,\quad} l }
    \frac{3^k}{e^3 + 1} + O(\sigma^{-\frac{1}{2}}),
       & {\rm for \; N=1,} \\
    3^k + O(\sigma^{-\frac{1}{2}})& {\rm for \; N\geq 2.}
   \end{array} \right.
\]
\item{$\underline{\alpha < -3}$:}
\[
   \stackrel{\circ}{v}_{2k}\; = \;
   \cA_{N,k} \; ( 1-\frac{2\alpha}{3})^k \;
   + \; O(\sigma^{-\frac{1}{2}}).
\]
\end{description}
\end{lemma}

For $\alpha>1$ the vertices are identical with those of the
O(N) Heisenberg models. In the range $-3<\alpha<1$
they agree with the vertices of a purely Gaussian model with ultralocal
action
\[
  \stackrel{\circ}{S}(\Phi) = (3-2\alpha)\sigma \Phi^2,
\]
leading to complete disorder as $\sigma\to\infty$
for every finite $\kappa$.
For $\alpha<-3$ we obtain the Heisenberg model again, as can be
seen as follows. For any $\beta>0$, rescaling of the vertices
\be \label{app1.scale}
\stackrel{\circ}{v}_{2k} \to \beta^{2k}\stackrel{\circ}{v}_{2k}
\ee
implies a corresponding rescaling of the connected vertices,
$\stackrel{\circ}{v}_{2k}^c \to \beta^{2k}\stackrel{\circ}{v}_{2k}^c$,
cf.~\eqn{lce.vcc}.
In turn, elementary graph theory shows that
all susceptibilities change according to
$\chi_n(\kappa)\to\beta^n \chi_n(\beta^2\kappa)$.
Hence, universality classes are invariant under \eqn{app1.scale}.
Furthermore we see that, for $N\geq 2$,
the boundary points $\alpha=1$ and $\alpha=-3$
belong to the Heisenberg class as well.
A remnant of the occupation number variables is only seen in the
case of $N=1$, which is a remarkable exception and needs further study.

Thermodynamic quantities like $\chi$ and the critical
coupling $\kappa_c$ can be directly determined in these limiting
models. Alternatively, one may start with the original action
\eqn{app1.actal} at finite $\sigma$,
calculate $\chi$ and $\kappa_c$ in the HPE, and take the large
coupling limit last. Our results indicate that both limits commute.


\end{appendix}


%
%


\end{document}